\documentclass{article}

\usepackage{arxiv}

\usepackage[utf8]{inputenc} 
\usepackage[T1]{fontenc}    
\usepackage{hyperref}       
\usepackage{url}            
\usepackage{booktabs}       
\usepackage{amsfonts}       
\usepackage{nicefrac}       
\usepackage{microtype}      
\usepackage{lipsum}
\usepackage[font=small,labelfont=bf]{caption} 

\usepackage{graphicx,calc}
\usepackage{amsmath}
\usepackage[english]{babel}
\usepackage{eurosym}
\usepackage{stmaryrd}

\usepackage{comment}
\usepackage{stmaryrd}
\usepackage{cite}
\usepackage[table,xcdraw]{xcolor}
\usepackage{lineno}

\graphicspath{{IMGs//}}

\title{Towards predicting shear-banding instabilities in lipid monolayers}

\author{
  A.R. Carotenuto \\
  Department of Structures for Engineering and Architecture, University of Napoli "Federico II" - Italy\\
  \texttt{angelorosario.carotenuto@unina.it} \\
  \And
  A. Gaffney \\
  Department of Surgery, Section of Vascular Surgery and Endovascular Therapy, University of Chicago (IL) - USA\\
  \texttt{adgaffney@uchicago.edu}\\
  \And
  N. Nguyen\\
   Department of Surgery, Section of Vascular Surgery and Endovascular Therapy, University of Chicago (IL) - USA\\
   \texttt{nhungng@bsd.uchicago.edu}\\
   \And
  K. Y. C. Lee\\
   Department of Chemistry and Institute for Biophysical Dynamics, University of Chicago (IL) - USA\\
   \texttt{kayeelee@uchicago.edu}\\
   \And
 L. Pocivavsek\\
   Department of Surgery, Section of Vascular Surgery and Endovascular Therapy, University of Chicago (IL) - USA\\
   \texttt{lpocivavsek@bsd.uchicago.edu}\\
   \And 
  M. Fraldi\\
  Department of Structures for Engineering and Architecture, University of Napoli Federico II, Italy\\
  Institute for Applied Sciences and Intelligent Systems, National Research Council of Italy\\
  \texttt{fraldi@unina.it}\\
  \And
  L. Deseri\\
  Department of Civil, Environmental and Mechanical Engineering, University of Trento - Italy\\
  Department of Mechanical Engineering and Material Sciences, SSoE, University of Pittsburgh USA\\
  Department of Civil and Environmental Engineering, Department of Mechanical Engineering, Carnegie Mellon, USA\\
  Department of Nanomedicine, The Houston Methodist Research Institute, USA\\
  \texttt{luca.deseri@unitn.it}
}

\begin{document}
\maketitle

\begin{abstract}
Langmuir monolayers are advantageous systems used to investigate how lipid membranes get involved in the physiology of many living structures, such as collapse phenomena in alveolar structures. Much work focuses on characterizing the pressure-bearing capacity of Langmuir films, expressed in the form of isotherm curves. These show that monolayers experience different phases during compression with an according evolution of their mechanical response, incurring into instability events when a critical stress threshold is overcome.
Although well-known state equations, which establish an inverse relationship between surface pressure and area change, are able to properly describe monolayer behaviour during liquid expanded phase, the modelling of their nonlinear behaviour in the subsequent condensed region is still an open issue. In this regard, most efforts are addressed to explain out-of-plane collapse by modelling buckling and wrinkling mainly resorting to linearly elastic plate theory.
However, some experiments on Langmuir monolayers also show in-plane instability phenomena leading to the formation of the so-called shear bands and, to date, no theoretical description of the onset of shear banding bifurcation in monolayers has been yet provided. For this reason, by adopting a macroscopic description, we here study material stability of the lipid monolayers and exploit an incremental approach to find the conditions that kindle shear bands. In particular, by starting from the widely assumed hypothesis that monolayers behave elastically in the solid-like region, in this work a hyperfoam hyperelastic potential is introduced as a new constitutive strategy to trace back the nonlinear response of monolayer response during densification. In this way, the obtained mechanical properties together with the adopted strain energy are successfully employed to reproduce the onset of shear banding exhibited by some lipid systems under different chemical and thermal conditions.
\end{abstract}

\section{Introduction}
\label{intro}

\begin{figure}[htbp] 
\centering
\includegraphics[width=.99
 \textwidth]{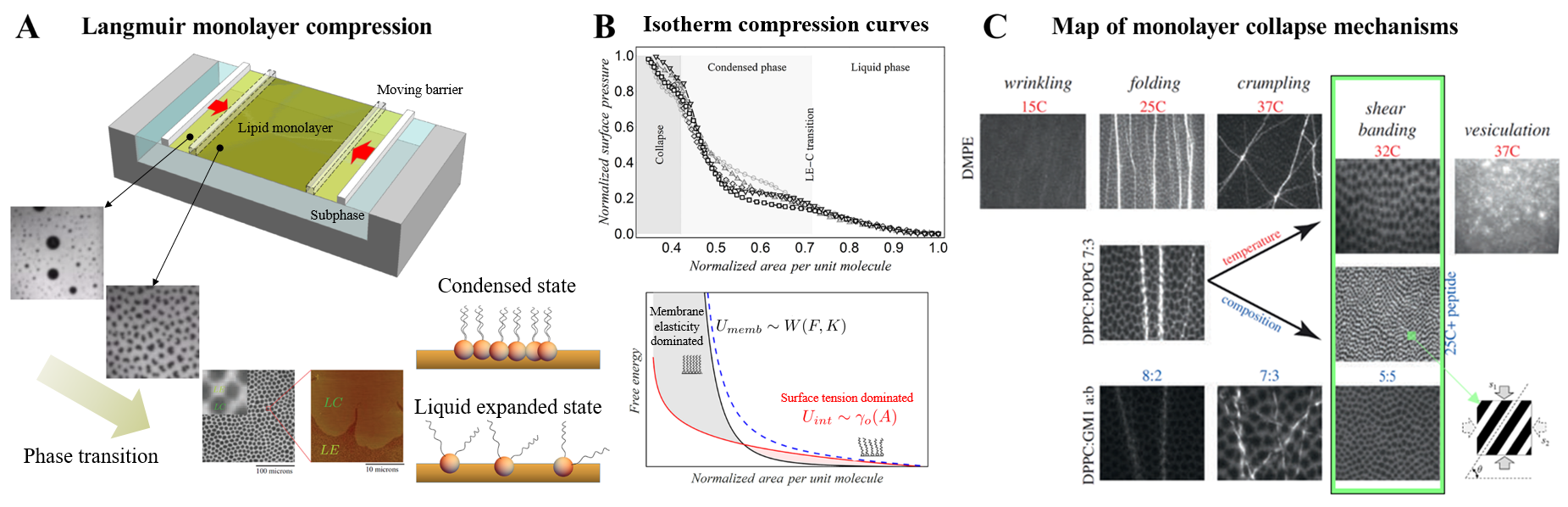}
\caption{\textbf{A} Compression test of a lipid monolayers in a Langmuir though and formation of monolayer patterns during transitions from liquid expanded to condensed phases. \textbf{B} (\textit{Top}) Examples of isotherm curves for different monolayers, which show different responses at the onset of condensed region.(\textit{Bottom}) Qualitative representation of lipid monolayer energy as a function of the normalized area per lipid molecule. At low packing density (gas and LE phases), the monolayer energy is dominated by the interfacial energy. As the monolayer transitions into an elastic solid, beginning with the LC phase, a membrane elastic energy term emerges, with a rigidity that governs the collapse mechanisms. \textbf{C} Generalized monolayer collapse phase diagram showing the richness of collapse states in lipid monolayers and their tunability with various field parameters including composition and temperature. Monolayer collapse is an elastic instability whereby the lipid membrane collectively over some length scale undergoes in-plane or out-of-plane displacement. This is a fundamentally different mode than movement of small clusters of lipids from the surface to the bulk phase well described by Gibbs’ absorption monolayers. The collective collapse modes are inherently governed by membrane elasticity (adapted from \cite{carotenuto2021multiscale}).}
\label{fig.fig1}
\end{figure}

Langmuir monolayers provide rich model systems for studying the behaviour of biological membranes \cite{Mohwald, Colloids}.  These systems are often employed to investigate the complex interactions between cell membranes and proteins/polymers \cite{Gidalevitz, Wu}, or inter-membrane interactions between lipids and various molecules such as cholesterol \cite{Ege}. Besides lipid bilayers, a key building block for cells, several biological membranes are actually mono-layered structures, the most prominent one being lung surfactant \cite{Zasadzinski}. 
Many aspects of lipid membranes are investigated in depth for their key importance in maintaining the homeostasis of cells and organs. Emphasis is put on the chemical interactions at play, i.e. electrostatic interaction between a charged lipid and charged peptide (the packing structure of lipid tails), as well as on some mechano-biological phenomena at larger scales involving the membrane re-organization and stress triggered by the activation of trans-membrane proteins\cite{carotenuto2020raft}. Indeed, the more coarse grained mechanical properties of the membrane play an equally important role in mediating biological functions \cite{Boal,Fung}. \textcolor{black}{For instance, lung surfactants are monolayer structures spread in alveoli composed of about 90$\%$ lipids and 10$\%$  surfactant-specific proteins (SP), namely SP-A , SP-B, SP-C and SP-D, which are known to be essential in structural response and in some immunoregulatory functions \cite{haagsman2001surfactant, wright2005immunoregulatory, garcia2019structure}. In particular, SP-A and SP-D are members of a family of immune proteins known as collectins that interact with various pathogens and enhance their uptake by phagocytes, by regulating functions of a variety of immune cells, including dendritic cells, T cells, neutrophils and macrophages. On the other hand, SP-B is fundamental for the correct activity of the pulmonary surfactant complex being it also involved into the process of surfactant membrane biogenesis, while SP-C contributes to stabilise surfactants films during respiratory cycle.
From a structural standpoint, pulmonary surfactants are so essential interfaces to prevent alveolar collapse, so acting as a true surfactant, lowering the surface tension by reducing surface tension at the pulmonary air–liquid interface and, consequently, lowering the hoop stress experienced by alveoli\cite{bernhard2016lung}. Such stress would otherwise undergo severe Laplace pressures due to their small dimensions.}
If forced to work against the high surface energy of its huge air-water interface, lung tissues can become mechanically unstable, thereby increasing their chances of collapse. However, in stabilizing the lungs against mechanical collapse, lung surfactant is itself put under high compressive stresses. Indeed, as the alveolar surface area undergoes very large expansions and shrinking during breathing, the surfactant layer experiences cyclic compression and decompression accordingly. Under such compression, the thin surfactant layer becomes denser than in the decompression phase. When a fully compacted state is reached such a layer becomes mechanically unstable under compression, as it begins to deform out of plane from its locally flat geometry \cite{Lipp, Gopal, Ybert}.
Nonetheless, these latter geometrical changes are not the only manifestation of instabilities in lipid monolayers.
Investigations towards this direction may be relevant to \textit{Acute Respiratory  Distress Syndromes}, ARDS, where lung surfactant instabilities seem to be a key phenomenon occurring in such an impairing illness.  
This is due to the fact that many times surfactants behave as granular bodies \cite{Pocivavsek1, Chachaj-Brekiesz2019}, thereby owing in-plane rearragements of lipid clusters.
\color{black}
From the view point of modelling such phenomenon, in \cite{Pocivavsek1} it is stated that the surfactant under stress must collapse in some mechanically reversible way. Besides maintaining the mechanical stability of the entire lung, reversibility assures that the surfactant covered layer remains intact throughout the breathing cycle and so preserving its protecting function against lung tissue collapse. 
Recent modeling explored potential instabilities occuring at the nanoscale  \cite{Tempra}. Typical sizes of the explored prototypical regions were of the order of 100 $\mathring{\mathrm{A}}^2$, hence far from detecting mesoscopic bifurcating events such as shear banding. 
\paragraph{Mechanical response and buckling of lipid monolayers.}
With the aim to analyze the leading aspects driving these instability mechanisms, Langmuir monolayers have proven to be ideal systems to study lipid mechanics in either highly confined geometries or high lipid packing densities. These systems can be in fact compressed or expanded by modifying their area with a moving barrier in a Langmuir film balance. Such hard-device experiments allow for obtaining isotherm curves relating the variation of surface pressure with the area (per unit molecule) change (see Figure \ref{fig.fig1}\hyperref[{fig.fig1}]{A}). 
Lipid monolayers are complex materials with accordingly complex mechanical responses. Indeed, they exhibit a rich variety of behaviours by progressively passing through different phases, from gaseous to condensed (solid-like) ones. 
Similarly to many microstructured materials, for two-component Langmuir monolayers the response to lateral squeezing is expected to directly influence its overall mechanical properties and evolves as a function of the conformational changes shown by these particular systems. More in detail, \textcolor{black}{during the first stages of compression, long range lateral interactions and water-lipid interactions are mostly involved with negligible gel-like/solid-like interactions.} Herein, the lipid molecules undergo phase transitions from the low density \textit{gas} phase to a more condensed \textit{liquid expanded} phase (LE). In these stages, the response of the system is well described by isotherms pressure curves establishing an inverse relationship between pressure and area change, as also qualitatively represented in Figure \ref{fig.fig1}\hyperref[{fig.fig1}]{B}. In particular, the area available to the monolayer $A$ (reciprocal of the surface density) decreases as the surface pressure $\pi$ rises by lowering the normal air-water surface tension from $\gamma_0$ to $\gamma$, i.e. $\pi=\gamma_0-\gamma$ \cite{Lipp,Pocivavsek1,Pocivavsek2,Pocivavsek3,Pepicelli2017}. If compression in liquid expanded phase can be described through firm thermodynamic arguments, less efforts are to date made to model the macroscopic behaviour of monolayers during the solid phase. Above a certain level of imposed squeezing, a second first order phase transition indeed occurs when \textit{liquid condensed} (C) domains nucleate and grow from the liquid phase. From a macroscopic point of view, at the beginning of condensation the pressure-area curves of monolayers show an initial tract that can exhibit a great variety of mechanical responses (see Figure \ref{fig.fig1}\hyperref[{fig.fig1}]{B}). They are characterized by either quasi-horizontal plateaus at almost constant stress and low modulus --typically associated to pure phase transitions-- or progressively increasing bulk rigidities, during which the system exhibits higher tangent slopes. Such differences mainly depend on the composition of the mixture and on temperature conditions. This implies that the starting of the condensed phase is due to variable mechanisms by which lipid clusters start to interact each other, by producing different overall bulk rigidities.
Once nucleation and condensation of the micro-domains has occurred, the monolayer experiences a subsequent densification phase (with more common characteristics among the several lipid systems), where the domains tends to compact and pack together, resulting in a rapid increase of compressive stress. 
In this solid region, it is a well-established hypothesis that Langmuir monolayers are believed to behave almost like an elastic solid, leading to explain how exceeding compressive stress may lead to the monolayer collapse \cite{ries1979stable, milner1989buckling,Pocivavsek1}. This event generally involves the interplay of the accumulated stresses with the extensional and bending rigidities of the lipid film. In general, such features are also sensitive to either material imperfections \cite{cutolo2022class} (which can be here given in this case by the presence of surfactant molecules) or the action of small driving forces (such as gravity effects \cite{milner1989buckling}).\\
As summarized in Figure \ref{fig.fig1}\hyperref[{fig.fig1}]{C}, experiments on Langmuir monolayers show that different collapse mechanisms may take place, mainly influenced by monolayer composition, surface density, sub-phase and temperature conditions. Out-of-plane collapse modes are the most common ones, in which monolayers behave as plates under compression by either undergoing mechanical buckling and loosing their flat configuration or by forming wrinkles that evolve into folds or creases made visible as brighter detaching strips at fluorescence imaging \cite{Pocivavsek3, Lipp,lee2008collapse, leahy2010geometric}. 
Buckling and wrinkling-to-folding transitions represent collapse mechanisms that have been extensively studied in lipid monolayers both experimentally and theoretically, by means of analytical and numerical techniques. The latter are based on known continuum mechanics methods. In this same category, it is worth noting that nonlinear elasticity has been recently utilized to investigate the response of lipid monolayers under compression \cite{Pepicelli2017}, without regards to any instabilities that such systems may exhibit.

Experimental work on model lung surfactant mixtures using Langmuir troughs has provided great insight into the collapse mechanism of these thin films. Much of the work has focused on two-component lipid monolayers using a saturated and unsaturated lipid, with and without the surfactant peptides. The mode of collapse initially observed in \cite{Lipp} showed the co-existence of buckled and flat monolayers in systems containing lung surfactant protein B and only one lipid component. Later, in \cite{Gopal} a similar transition in a two component lipid system excluding the peptide was shown, thus proving that lipids alone could support a reversible collapse mode.  In both cases, the monolayer is ina fully compact state and forms well defined macroscopic folds perpendicular to the direction of compression. These folds form only at very high compressions corresponding to near zero surface tensions at the air/lipid/water interface. The folds are reversible such that upon decompression one can visualize their unfolding and subsequent re-spreading of the monolayer onto the surface.  
The origin of the folding instability in these biphasic monolayers has received much attention. In particular, theoretical analysis performed in \cite{Diamant1, Diamant2} focused on the intrinsic mismatch in elastic properties that exists in a biphasic system. They showed that at the interface between two phases in a monolayer an instability develops at high compression.  This instability takes the form of overhangs or so called mesas structures that could provide the nucleation sites for the macroscopic folds. 
From a modelling standpoint, well established approaches based on Helfrich-type curvature effects provided the buckling analyses of monolayers treated as elastic sheets obeying beam equation under linear constitutive and geometric assumptions \cite{ries1979stable, milner1989buckling, vella2004elasticity, Pocivavsek1, Pocivavsek2, leahy2010geometric}, by providing estimations of critical pressures and undulations wavelengths at the onset of buckling. Augmented Helfrich-like approaches (currently available for lipid bilayers only) would allow for following changes in bending (and gaussian) stiffness due to squeezing and compaction \cite{Deseri:2008, Deseri:2013, Pollaci:3}. The analysis of stability of these systems could be improved with the help of structural paradigms that include mechanical nonlinearities in order to study the effect of macroscopic prestretch as well as the influence of sub-macroscopic interactions appropriately translated as elastic connections between lipid elements \cite{palumbo2018nonlinear, palumbo2021bulky, fraldi2021equilibrium}. Molecular dynamics investigations have been proposed to simulate the evolution of initial buckling instabilities into folding and possible formation of vesicles \cite{baoukina2008molecular}, while other approaches enrich classical elastic modelling assumptions by including granular effects of rafts \cite{cicuta2009granular}.\\ 

\paragraph{Banding as additional in-plane instability.} Although the vast majority of the above mentioned work focuses on out-of-plane instabilities, some experimental evidences have instead shown the occurrence of in-plane instability mechanisms. This event has been mostly attributed to different mechanical properties of the monolayers deriving from to the composition of raft forming membranes. In \cite{Pocivavsek2} it is indeed shown that monolayers that can be pushed through a fully compacted state show a high rigidity during densification (i.e. during the last hardening curve of the condensed phase while looking at the corresponding isotherm), hence the only way they have to relax their stress is to collapse out-of-plane. On the other hand, monolayers exhibiting a more granular behavior, hence softer than the ones just discussed above, can be prone to relax their stress through in-plane shear banding. The latter instabilities are characterized by localization of the deformation accompanied by an in-plane rearrangement of lipid domains. Accordingly, by adopting a viscoelastic approach, in \cite{majumdar2011shear} shear bands are viewed as the inhomogeneities of strain rate manifested in a series of 
experiments. Such bands in \cite{majumdar2011shear} were interpreted by appealing to a ''plastic-like flow'' with loss of rigidity and stress plateau at instability. 
\textcolor{black}{The formation of shear bands at $\pm 45^\circ$ with respect to the stretching direction has been observed in different literature experiments on lipid systems, describing monolayers subjected to extensional flow with or without the combination of rotational fields \cite{maruyama1998orientation, maruyama1996flow} as well as shear-induced distortion and reorientation of monolayers' hexatic domains \cite{iwamoto2008shear, ignes2001shear}.}
As it is well known, shear bands can be viewed as regions of highly concentrated deformation developing from slow varying strain state. Continuum mechanics allows for analyzing such bands as preferential near-failure deformation modes of many ductile materials \cite{Rice76, bigoni2008unrestrainable} as well as hyperelastic solids (see e.g. \cite{bigonicapuani2002,bigoni2012nonlinear}). For this reason, the derivation of the conditions for the existence of a shear band has been largely investigated in the framework of $J_2$-deformation theory of plasticity \cite{hutchinson1981finite, hutchinson1981shear, mengi1975theory,  bigoni2012nonlinear}, mainly with reference to metals and granular media. In all these papers, shear banding are treated in the framework of material instability.
At the microscale level, partially condensed lipid systems seem to exhibit many similarities with granular media, by showing packing of the single lipid clusters as well as positional disarrangements involving relative switching and reorientation similar to jamming \cite{carotenuto2021multiscale}. These could be included through an enriched multi-scale geometry based on Structured Deformation theory \cite{carotenuto2021multiscale, palumbo2021bulky}. Describing the macroscopic response of the lipid monolayers --say at the level of a patch of packing domains of the order of tens/hundreds of microns-- would involve a characterization of the disarrangements arising at that level and, hence, understanding their (possibly dissipative) evolution.
A simplified approach to gain first insights into deriving the conditions for the formation of shear bands in lipid monolayers is to assume (i) a macroscopic, one scale, kinematics and (ii) they behave like compressible hyperelastic materials at finite strains while undergoing an overall imposed contracting deformation. It is worth noting that the emergence of shear bands in purely nonlinear elastic materials has been examined with reference to different choices of elastic energies \cite{reddy1983occurrence, abeyaratne1981emergence, knowles1978failure, hill1975bifurcation, hutchinson1981shear, tvergaard1981flow, kurashige1983shear, gultop2002existence, bigoni2012nonlinear}. 
Under these assumptions, the purpose of the present paper is to explain mechanically the potential occurrence of shear banding observed in some experiments, by suitably modelling the macroscopic behaviour of the Langmuir monolayers through the introduction of appropriate constitutive assumptions. They will have to be able to trace back the finite response of the different monolayers during the entire condensation and compaction (solid) phase.
As discussed above, the condensed phase exhibits fickle responses during transition and compaction, with different evolving properties. Consequently, the need of reproducing such a variety of behaviours in these stages requires the adoption of suitable material models. The latter should have the capability to provide both variation of density (through material compressibility) and the possibility of reproducing more complex features with respect to classical hyperbolic-type responses. Such features would relate to ad hoc parameters shaping the curve through the different stages of condensation. Because of all these aspects qualitatively resembling micro-structural response of cellular solids, in what follows we propose that the \textit{hyperfoam} constitutive model represents a good candidate to describe the elastic responses of the monolayers at the macroscopic scale \cite{storaakers1986material}. Indeed, this particular constitutive law has the capability to simulate the possible nonlinear trends exhibited by isotherm curves in the C phase. We then employ this material model to find effective properties of monolayers in the solid-like region.
These outcomes are then used to mathematically estimate the emergence of shear bands during compression. To do this, we make use of an incremental approach in order to derive the general conditions that govern the loss of ellipticity (l.o.e.) of the tangent stiffness tensor. This leads an homogeneously deformed representative area element of the monolayer to exhibit solutions in terms of the displacement fields in the hyperbolic regime, so carrying characteristic discontinuity lines \cite{hutchinson1981shear, biot1965mechanics, bigoni2012nonlinear, kurashige1983shear} that produce shear bands. The l.o.e. conditions are then particularized to the hyperfoam material law with the derived parameters. This allows for reproducing banding instabilities occurring in some experiments performed on Langmuir monolayers. In this way an estimation of the potential banding pressure and of the band slopes at incipient collapse is provided in direct comparison with experimental evidences. In particular, the proposed theoretical framework will allow for predicting if and when banding occurs by testing the capability of this mathematical approach against experimental evidences from different lipid systems.\\

\section{Conditions for shear banding in hyperelastic media}
\label{model}

\subsection{Tangent moduli and incremental constitutive equations}
In order to analyse the possible shear-band bifurcation occurring in lipid monolayers under finite deformations, we assume that the lipid system subjected to monotone isothermal compression in the solid (condensed) phase macroscopically behaves as a hyperelastic material. Starting from this assumption, we evaluate the conditions by which a finite and homogeneous deformation state of a small region of the two-dimensional monolayer is able to kindle in-plane instability. To do this, we lean on Biot's theory of incremental deformations \cite{biot1965mechanics} in order to explore the conditions for which the incremental equilibrium equations exhibit a loss of ellipticity, the corresponding displacement solutions providing a possible discontinuity in their first derivative, which is known as elasto-static shock line \cite{knowles1980discontinuous}. 
Formally, the problem is presented by considering the presence of possible disarrangements, which enrich the classical kinematics and take into account the possible presence of positional and conformational changes of lipids occurring at the sub-macroscopic scale. These are modelled according to the Structured Deformation theory \cite{DelPiero1993,deseri:2003,Deseri:2010,deseri:2019,deseri:2012, palumbo:2018, palumbo2017nonlinear} and represent additional terms of deformations associated to lipid microstructural re-arrangements. These are projected at the continuum scale by means of a so called disarrangement tensor, obtained through limit procedures \cite{carotenuto2021multiscale}. The potential presence of these terms could make possible to separate the macroscopic (classical) deformation into a disarrangement-free quota of the deformation gradient (characterized by limits of bulk averages of gradients away from disarrangement sites \cite{carotenuto2021multiscale, DelPiero1993,deseri:2003,Deseri:2010,deseri:2019,deseri:2012, palumbo:2018}) and a contribution due to sub-macroscopic disarrangements, so basically discriminating the elastic and conformational quotas of the stretch. 
In what follows, the first analyses will however consider the sole effect of canonical, namely macroscopic, deformations by \textit{de facto} not distinguishing the contributions mentioned above. Henceforth, in the sequel the disarrangement-free deformation measure, $\mathbf{G}$, 
coincides with the usual macroscopic deformation gradient $\mathbf{F}$.
It follows that the reference (Piola-Kirchhoff) stress, $\mathbf{P}_0$, is given by
$\mathbf{P}_0=\,\partial\, W_0(\mathbf{G})/\partial \mathbf{G}$, where 
$W_0$ indicates 
the hyperelastic strain energy density in the natural configuration, depending on $\mathbf{G} = \mathbf{F}$ though its invariants. 
The \textcolor{black}{rate of change} of the corresponding Cauchy stress \textcolor{black}{$\boldsymbol{\sigma}_0$} can be expressed as:

\begin{equation}\label{incrCauchy1}
\dot{\boldsymbol{\sigma}}_0=\dot{\left[\overline{\frac{1}{J_G}\,\mathbf{G}\,\mathbf{S}_0\,\mathbf{G}^T}\right]}=-tr(\mathbf{L}_G)\boldsymbol{\sigma}_0+\mathbf{L}_G\boldsymbol{\sigma}_0+\boldsymbol{\sigma}_0\mathbf{L}_G^T+\frac{1}{J_G}\,\mathbf{G}\,\dot{\mathbf{S}}_0\,\mathbf{G}
\end{equation}

where $\mathbf{L}_G=\dot{\mathbf{G}}\mathbf{G}^{-1}$ is the 
velocity gradient, $\mathbf{S}_0=\mathbf{G}^{-1}\mathbf{P}_0$ is the second Piola-Kirchhoff stress and $J_G=\det \mathbf{G}$.
By keeping the 
natural configuration fixed 
the increment of $\mathbf{S}_0$ reads:

\begin{equation}\label{incr2PK}
\dot{\mathbf{S}}_0=2\frac{\partial^2 W_0(\mathbf{G})}{\partial \mathbf{C}_G\otimes\partial\mathbf{C}_G}:\dot{\mathbf{C}}_G=2\mathbb{G}:\dot{\mathbf{C}}_G
\end{equation}

where $\mathbf{C}_G=\mathbf{G}^T\mathbf{G}$ is the right Cauchy-Green strain tensor associated to $\mathbf{G}$ conjugated to $\mathbf{S}_0$, \textcolor{black}{while the $ijhk$-th component of the fourth-order tensor $\mathbb{G}$ is obtained as $\text{G}_{ijhk}=\partial^2\,W_0/\partial\,C_{G\,ij}\partial\,C_{G\,hk}$ and the column denotes  contraction of two indexes, i.e. $\left[\mathbb{A}:\textbf{B}\right]_{ij}=A_{ijmn}B_{mn}$}. By substituting \eqref{incr2PK} into \eqref{incrCauchy1}, after some steps one obtains:


\begin{equation}\label{incrCauchy2}
\dot{\boldsymbol{\sigma}}_0=-tr(\mathbf{L}_G)\boldsymbol{\sigma}_0+\mathbf{L}_G\boldsymbol{\sigma}_0+\boldsymbol{\sigma}_0\mathbf{L}_G^T+\frac{4}{J_G}\,\left(\mathbf{G}\overline{\otimes}\mathbf{G}\right):\mathbb{G}:\left(\mathbf{G}^T\overline{\otimes}\mathbf{G}^T\right):\mathbb{S}:\mathbf{L}_G,
\end{equation}

in which $2\mathbb{S}=\left(\mathbf{I}\overline{\otimes}\mathbf{I}+\mathbf{I}\underline{\otimes}\mathbf{I}\right)$ denotes the identity fourth-order tensor for symmetric tensors \textcolor{black}{($\mathbf{I}$ being the second-order identity tensor)}, while the not standard tensor products are defined as $\left[\mathbf{A}\overline{\otimes}\mathbf{B}\right]_{ijhk}=A_{ih}B_{jk}$ and $\left[\mathbf{A}\underline{\otimes}\mathbf{B}\right]_{ijhk}=A_{ik}B_{jh}$. In this way, by considering that $\mathbf{L}_G$ can be decomposed in its symmetric and skew-symmetric parts, respectively denoted as $2\mathbf{d}_G=\mathbf{L}_G+\mathbf{L}_G^T$ and $2\mathbf{w}_G=\mathbf{L}_G-\mathbf{L}_G^T$, relation \eqref{incrCauchy2} can be written in terms of Jaumann co-rotational rate of the Cauchy stress, defined as:

\begin{equation}\label{incrCauchy3}
\mathring{\boldsymbol{\sigma}}_0:=\dot{\boldsymbol{\sigma}}_0-(\mathbf{w}_G\boldsymbol{\sigma}_0-\boldsymbol{\sigma}_0\mathbf{w}_G^T)=\mathbb{M}_0:\mathbf{d}_G
\end{equation}

where 


\begin{equation}\label{M0}
\mathbb{M}_0=-\boldsymbol{\sigma}_0\otimes\mathbf{I}+\mathbf{I}\overline{\otimes}\boldsymbol{\sigma}_0+\boldsymbol{\sigma}_0\underline{\otimes}\mathbf{I}+\frac{4}{J_G}\,\left(\mathbf{G}\overline{\otimes}\mathbf{G}\right):\mathbb{G}:\left(\mathbf{G}^T\overline{\otimes}\mathbf{G}^T\right)
\end{equation}

is the tangent fourth-order tensor referred to the virgin configuration, which provides the presence of internal pre-stress in a linear fashion and of tangent moduli nonlinearly depending on finite stretches. Because here $\mathbf{G} \equiv \mathbf{F}$, it follows that 
%
$\mathbf{L}_G=\mathbf{L}$, hence $\mathbf{d}_G = \mathbf{d}$.
%
Here we write the incremental problem by starting from a finitely deformed configuration with homogeneous deformations and stresses. By thus denoting the incremental stress as $\mathbf{s}=\mathring{\boldsymbol{\sigma}}_0\,dt$ and the incremental strain as $\boldsymbol{\epsilon}=\mathbf{d}\,dt$ in the Biot's formalism and by referring to a two-dimensional strain energy density $W_0(\mathbf{G})=W_0(I_G,J_G)=\tilde{W}_0(G_1,G_2)$, where $I_G=tr(\mathbf{C}_G)$, according to the above definitions, the stress $\boldsymbol{\sigma}_0$ is given by:


\begin{equation}\label{CauchyStress1}
\boldsymbol{\sigma}_0=\frac{1}{J_G}\frac{\partial {W}_0}{\partial \mathbf{G}}\mathbf{G}^T=\frac{1}{J_G}\left(\frac{\partial W_0}{\partial I_G}\frac{\partial I_G}{\partial \mathbf{G}}+\frac{\partial W_0}{\partial J_G}\frac{\partial J_G}{\partial \mathbf{G}}\right)=
2J_G^{-1}W_{0,I}\,\mathbf{b}_G +W_{0,J}\mathbf{I}=\frac{\tilde{W}_{0,i}}{G_j}\mathbf{n}_i\otimes\mathbf{n}_i, \quad i\neq j
\end{equation}

where $\mathbf{b}_G=\mathbf{G}\mathbf{G}^T$ is the left Cauchy-Green tensor and $\tilde{W}_{0,i}=\partial\,W_0/\partial\,G_i$. Also it results:

\begin{equation}
\frac{\partial {W}_0}{\partial \mathbf{C}_G}= W_{0,I} \mathbf{I}+W_{0,J}\frac{J_G}{2}\mathbf{C}_G^{-1}
\end{equation}

and

\begin{eqnarray}\label{Gw}
\mathbb{G}&= W_{0,II}\, \mathbf{I}\otimes\mathbf{I}+\frac{J_G}{2}\left(W_{0,IJ}\,\mathbf{C}_G^{-1}\otimes\mathbf{I}+W_{0,JI}\mathbf{I}\otimes\mathbf{C}_G^{-1}\right)+W_{0,JJ}\,\frac{J_G^2}{4}\mathbf{C}_G^{-1}\otimes\mathbf{C}_G^{-1}+\notag\\
&+W_{0,J}\,\frac{J_G}{4}\left(\mathbf{C}_G^{-1}\otimes\mathbf{C}_G^{-1}-\mathbf{C}_G^{-1}\overline{\otimes}\mathbf{C}_G^{-1}-\mathbf{C}_G^{-1}\underline{\otimes}\mathbf{C}_G^{-1}\right)
\end{eqnarray}

The use of relations \eqref{CauchyStress1} and \eqref{Gw} into \eqref{M0} let to find the following expression for the incremental stress:

%

\begin{eqnarray}\label{s}
\mathbf{s}&=\left\{\frac{2}{J_G}W_{0,I}\left[(\boldsymbol{\epsilon}\mathbf{b}_G+\mathbf{b}_G\boldsymbol{\epsilon}-tr(\boldsymbol{\epsilon})\,\mathbf{b}_G\right]+2W_{0,IJ}\left[tr(\boldsymbol{\epsilon})\,\mathbf{b}_G+tr(\boldsymbol{\epsilon}\,\mathbf{b}_G)\mathbf{I}\right]+\right.\notag\\
&+\left.\frac{4}{J_G}W_{0,II}tr(\boldsymbol{\epsilon}\,\mathbf{b}_G)\mathbf{b}_G+J_G\,W_{0,JJ}tr(\boldsymbol{\epsilon})\,\mathbf{I}\right\}
\end{eqnarray}

that is analogous to relation found for instance by Kurashige\cite{kurashige1983shear}. The incremental stress components then read as:

\begin{equation}
\begin{cases}
s_{11}=M^0_{1111}\,\epsilon_{11}+M^0_{1122}\,\epsilon_{22}\\
s_{22}=M^0_{2211}\,\epsilon_{11}+M^0_{2222}\,\epsilon_{22}\\
s_{12}=M^0_{1212}\,\epsilon_{12}
\end{cases}
\end{equation}

where the tangent moduli read as follows


\begin{equation}\label{moduliM}
\begin{cases}
M^0_{iiii}=\frac{G_i^2}{G}\tilde{W}_{0,ii}\\
M^0_{iijj}=\tilde{W}_{0,ij}-\frac{\tilde{W}_{0,i}}{G_j}, \quad i\neq j\\
M^0_{ijij}=\frac{I_G}{J_G}\left(\frac{\tilde{W}_{0,1}\,G_1-\tilde{W}_{0,2}\,G_2}{G_1^2-G_2^2}\right).
\end{cases}
\end{equation}

With reference to principal coordinates and by assuming a homogeneous finite deformation (with possibly homogeneous disarrangements) on which the incremental displacements are superimposed, the incremental equilibrium equations on a two-dimensional elementary area result \cite{biot1965mechanics}:

\begin{equation}\label{equilibrium1}
\begin{cases}
s_{11,1}+s_{12,2}-(\sigma_{01}-\sigma_{02})\omega_{12,2}=0\\
s_{12,1}+s_{22,2}-(\sigma_{01}-\sigma_{02})\omega_{12,1}=0
\end{cases}
\end{equation}

In the light of the expressions of tangent moduli \eqref{moduliM} and on account of the compatibility relations $\boldsymbol{\epsilon}+\mathbf{w}=\nabla\,\textbf{u}$, where $\textbf{u}=\{u_1,u_2\}$ is the in-plane incremental displacement in a way that $2\epsilon_{ij}=u_{i,j}+u_{j,i}$, the above equilibrium equations become:

\begin{equation}\label{equilibrium2}
\begin{cases}
\frac{G_1}{G_2}\,\tilde{W}_{0,11}\,u_{1,11}+\frac{G_2}{G_1}\,D_1\,u_{1,22}+D_2\,u_{2,12}=0\\[1.2ex]
\frac{G_2}{G_1}\,\tilde{W}_{0,22}\,u_{2,22}+\frac{G_1}{G_2}\,D_1\,u_{2,11}+D_2\,u_{1,12}=0
\end{cases}
\end{equation}

where 

\begin{align}
&D_1=\frac{G_1\,\tilde{W}_{0,1}-G_2\,\tilde{W}_{0,2}}{G_1^2-G_2^2}\notag\\
&D_2=\tilde{W}_{0,12}+\frac{G_2\,\tilde{W}_{0,1}-G_1\,\tilde{W}_{0,2}}{G_1^2-G_2^2}
\end{align}

\subsubsection{Classification of the incremental equations and shear-band bifurcation}

In order to provide the existence of possible discontinuities along characteristic lines with equation $c_1\,x_1+c_2\,x_2+\text{const.}=0$ in a homogeneously deformed representative area element (say a $\delta\times\delta$ RAE as the one studied in \cite{carotenuto2021multiscale}), the system of equations \eqref{equilibrium2} should result hyperbolic. 
By following Kurashige\cite{kurashige1983shear}, we thus consider displacement solutions in the form:
\begin{equation}\label{dispform}
u_1=f(c_1\,x_1+c_2\,x_2)\quad\text{and}\quad u_2=g(c_1\,x_1+c_2\,x_2)
\end{equation}

Substituting functions \eqref{dispform} in \eqref{equilibrium2} and by then requiring the existence of not trivial solutions under the validity of the convexity condition $\tilde{W}_{0,11}\,\tilde{W}_{0,22}>0$ with $\tilde{W}_{0,ii}\neq 0$ and $D_1\neq 0$, the following equation can be obtained:

\begin{equation}\label{ceq1}
\chi^2+2\,\rho\,\chi+\kappa^2=0
\end{equation}

where $\chi=c^2=c_1^2/c_2^2$, while the stretch-dependent parameters $\rho$ and $\kappa$ are defined through the relations: 
\begin{align}\label{positions}
\rho=\frac{E_{12}\,G_2^2}{E_{11}\,G_1^2} &\quad\text{and}\quad \kappa^2=\frac{E_{22}\,G_2^4}{E_{11}\,G_1^4},\notag\\
E_{ii}=&D_1\,\tilde{W}_{0,ii}\\
E_{12}=&\frac{1}{2}\left(\tilde{W}_{0,11}\tilde{W}_{0,22}+D_1^2-D_2^2\right)\notag
\end{align}

\begin{figure}[t] 
\centering
\includegraphics[width=.9
 \textwidth]{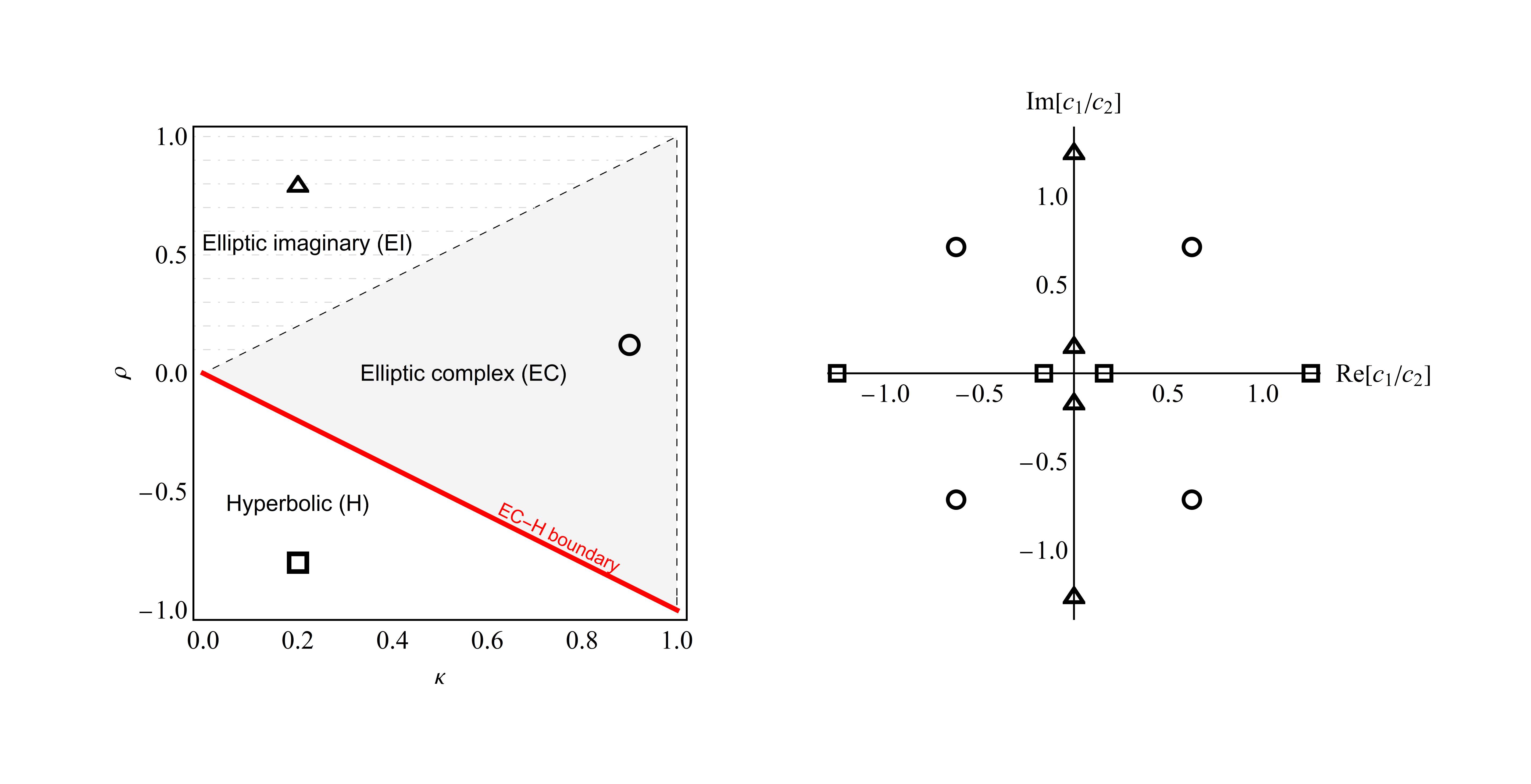}
\caption{Classification of the ratio $c_1/c_2$ as a function of the stretch-dependent parameters $\rho$ and $\kappa$, with identification of the region corresponding to real planes and associated to the insurgence of slip bands within the area element.}
\label{fig.class}
\end{figure}

Equation \eqref{ceq1} has solution $\chi=-\rho\pm\sqrt{\rho^2-\kappa^2}$. Therefore, the nature of the variable $c=\pm\sqrt{\chi}$ depends on the specific interaction of the parameters $\rho$ and $\kappa$. More precisely, provided that $\kappa>0$ under the above mentioned convexity requirement, the condition $\rho^2-\kappa^2<0$ determines the \textit{elliptic complex} (EC) region in which the $c$ ratio is complex, this implying that no real slip lines generate in the elementary domain. Similarly, the region $\{\rho>0,\,\rho^2-\kappa^2>0\}$ identifies \textit{imaginary} solutions (EI) for the variable $c$. On the other hand, the complementary domain $\{\rho<0,\,\rho^2-\kappa^2>0\}$  detects the \textit{hyperbolic region}(H), in which the ratio $c_1/c_2$ has real roots that return the equations of banding planes. This classification is highlighted in Figure \ref{fig.class}. For observing banding bifurcation, the boundary of interest is the one between the EC and the H regions, identified by the relationship $\rho=-\kappa$. In this limit situation, the solution to equation \eqref{ceq1} reduces to $\chi=\sqrt{\kappa}$, corresponding to the slopes

\begin{equation}\label{ratio}
c_0=\left.\frac{c_1}{c_2}\right|_{\text{EC-H}}=\pm\frac{G_2}{G_1}\,\left(\frac{\tilde{W}_{0,22}}{\tilde{W}_{0,11}}\right)^\frac{1}{4}
\end{equation}

and the incremental displacements solving the homogeneous system \eqref{equilibrium2} have expression:

\begin{align}\label{usol}
&u_1=f_{-}(x_2-c_0\,x_1)+f_{+}(x_2+c_0\,x_1),\notag\\
&u_2=\eta\left(f_{-}(x_2-c_0\,x_1)-f_{+}(x_2+c_0\,x_1)\right),
\end{align}

where the proportionality coefficient $\eta$ can be evaluated as

\begin{equation}
\eta=\frac{c_0^2\,\frac{G_1}{G_2}\,\tilde{W}_{0,11}+D_1\frac{G_2}{G_1}}{c_0\,D_2}
\end{equation}

while $f_{-}$ and $f_{+}$ are piecewise continuous functions carrying a discontinuity in their first derivative along the so-called elastostatic shock lines $L_\pm: x_2\pm\,c_0\,x_1+\text{const}=0$. The unit vectors normal to the shock lines will have expression $\mathbf{n}_{\pm}=\{\pm c_0,1\}/\sqrt{1+c_0^2}$, these allowing to calculate the continuity of the stress at the discontinuity lines by means of the interface condition:

\begin{equation}\label{traction}
\llbracket\, \left(\mathbf{s}+\mathbf{w}\,\boldsymbol{\sigma}_0+\boldsymbol{\sigma}_0\,tr(\mathbf{\boldsymbol{\epsilon}})-\boldsymbol{\sigma}_0\,\boldsymbol{{\epsilon}}\right)\cdot\mathbf{n}_{\pm}\rrbracket=\textbf{0}
\end{equation}

obtained by considering the time differentiation of the continuity balance for the reference tractions across the interface mapped back to the natural configuration, i.e. $\llbracket\,\mathbf{P}_0\cdot\mathbf{N}_0\,\rrbracket\,=\textbf{0}$. It can be easily verified that solutions \eqref{usol} automatically satisfy conditions \eqref{traction} for any $f_{\pm}$ continuous and piecewise smooth. Such a feature also ensures the continuity equations in terms of incremental displacements  $\llbracket\, u_i \,\rrbracket=0$ (no sum) at the discontinuity line.

\subsection{Hyperelastic potential for the lipid system}
The above relationships can be specialized by introducing a proper hyperelastic law for the lipid system in the condensed phase. It is worth noting that the phenomenological behavior of Langmuir interfaces has already been described like a nonlinear elastic material (see e.g. \cite{Pepicelli2017}), although neither their  explicit applicability to lipid monolayers have been analyzed nor investigations about potentially occurring material instabilities under compression have been performed.
As observed, the isotherms (also plotted in Figure \ref{fig.fig1}\hyperref[{fig.fig1}]{B})  \textit{de facto} provide data from which inferring pressure-area behaviour. 
As noticed, these curves all undergo typical hardening during the densification branch before exhibiting an instability point, while the first part after the LE-C transition instead shows a largely variable responses with different degree of compressibility. This might be ascribed to the particular packaging mechanisms and lipid interactions occurring at the micro-scale level. In turn, this influences the pressure responses at the macroscopic level by giving different compression moduli and starting curve shapes, which vary from either quasi horizontal plateaus preceding condensation or a more progressive hardening of the system during compression. 
This suggests to use hyperelastic laws able to phenomenologically reproduce curves with different macroscopic degree of compressibility. To this aim, we here adopt a Hyperfoam model (denoted as n-HM), with two-dimensional strain energy density given by the expression \cite{storaakers1986material, ogden1972large}:

\begin{equation}\label{SED}
W_n(G_1,G_2)=\sum_{i=1}^{n}\frac{E_i}{\alpha_i^2\,(1+\nu_i)}\left(G_1^{\alpha_i}+G_2^{\alpha_i}-2+\frac{1-2\nu_i}{\nu_i}\left((G_1 G_2)^{-\frac{\alpha_i\,\nu_i}{1-2\nu_i}}-1\right)\right)
\end{equation}

subject to the condition
\begin{equation}\label{ym}
\sum_{i=1}^{n}\frac{E_i}{2\,(1+\nu_i)}> 0.
\end{equation}

\begin{figure}[t] 
\centering
\includegraphics[width=.99
 \textwidth]{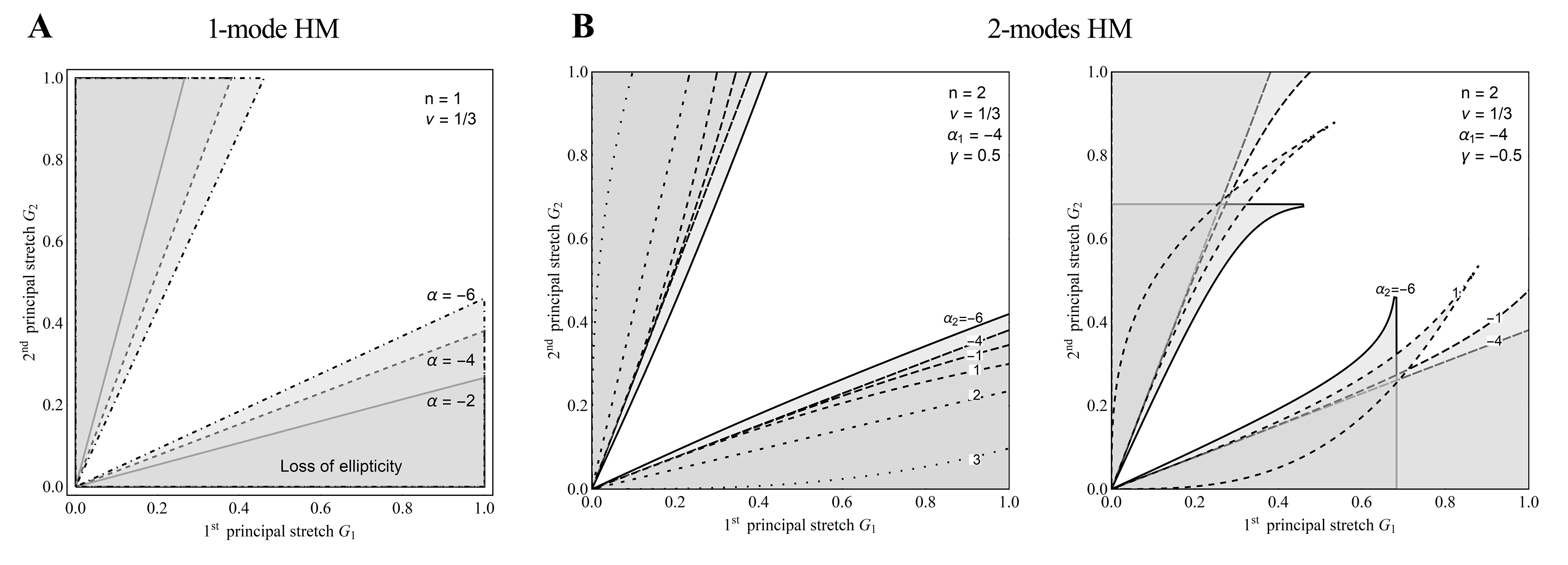}
\caption{Loss of ellipticity domains for different material constants in the HM. \textbf{A}. Influence of the parameter $\alpha$ in case of a 1-HM. \textbf{B}. Competition of the parameters $E_i$ and $\alpha_i$ in a 2-HM. Data: $\alpha=\alpha_1=-4$, $\nu=1/3$ and $\gamma=\pm 0.5$. }
\label{fig.loe}
\end{figure}

In this material model, common $\nu_i$ can be assumed for convenience to uniquely identify the characteristic Poisson coefficient, say $\nu$. This allows to determine from \eqref{ym} the initial Young modulus as the sum of the dimensional parameters $E_i$.
Also, the coefficients $\alpha_i$ represent Ogden-type parameters that work as shape factors by influencing the tangent moduli (as example, for $n=1$ and $\alpha=-2$ one recovers the Blatz-Ko compressible model). Consequently, the condition $\mathcal{L}(G_1,G_2)=\rho+\kappa=0$ determining the transition between the EC and H regime will depend on the stretch level mediated by the interplay of the adopted constitutive parameters. Importantly, this makes the loss of ellipticity (l.o.e.) not a pure geometric condition in which the critical stretch is not influenced by the physical moduli \cite{kurashige1983shear, hutchinson1981shear}. The obtained conditions for loss of ellipticity are explicitly reported in the \hyperref[{sec:appendix}]{Appendix}. Besides the dependence on specific values of $\nu$, they are affected by the parameter $\alpha$ for $n=1$ and, for $n=2$, by both the coefficients $\alpha_i$ and the ratio $\gamma=E_2/E_1$. Figure \ref{fig.loe} shows a sensitivity analysis of the l.o.e. domain --given by the condition $\rho+\kappa\leq 0$-- evaluated as a function of the principal stretches, for varying material constants. As highlighted, for a single-mode HM (Figure \ref{fig.loe}\hyperref[{fig.loe}]{A}), the extension of the l.o.e. domain in the contraction region of the $G_1-G_2$ plane varies with the amplitude of the coefficient $\alpha$, while it results to be not drastically modified by standard Poisson coefficients. A more complex interplay occurs in case of $n=2$, where multiple scenarios may take place by varying constitutive parameters. Figure \ref{fig.loe}\hyperref[{fig.loe}]{B} displays how the l.o.e. domain modifies for selected values of material constants. In particular, by using the same Poisson ratio, fixing $\alpha_1<0$ and assuming $E_1>0$, the competition of the stiffness-like parameters produces different alterations. With these positions, concord $E_i$ (i.e. $\gamma>0$) introduce more regular and progressive changes, the parameter $\alpha_2$ regulating the opening of the hyperbolic region, progressively extinguishing the l.o.e. domain when it increases. As illustrated in the right panel Figure \ref{fig.loe}\hyperref[{fig.loe}]{B},  $E_i$ opposite in sign (i.e. $\gamma<0$) and the difference between $\alpha_1$ and $\alpha_2$ can instead draw more drastic modifications of the shape of the l.o.e. sectors, consequently restricting the deformation conditions that could induce shear banding instability.
\\
In what follows, we apply this theoretical procedure to model the behaviour of different monolayer systems, by investigating the potential occurrence of shear banding in comparison to the actual experimental observations. To do this, we simplify our analysis by hypothesizing a uniaxial strain regime (by posing $G_2=1$, hence perfectly constraining the deformation in the transverse direction of loading) during compression, compatibly with the experimental conditions. This implies that the area change, the pressure and the l.o.e. condition can be written as a function of a single elastic stretch component that here coincides with the classical (macroscopic) stretch, say $G_1\equiv\lambda$ and $\mathcal{L}=\mathcal{L}(1,\lambda)$, in absence of strains due to disarrangements. 

\section{Application to lipid monolayers} 

\subsection{Experimental procedures}
\subsubsection{Lipids, Peptides, Subphase, Superphase}
DMPE (1,2-dimyristoyl-sn-glycero-3-phosphoethanolamine, $T_m=50^\circ\text{C}$),
DPPC (1,2-dipalmitoyl-sn-glycero-3-phosphocholine, $T_m=41^\circ\text{C}$), POPG
(1-palmitoyl-2-oleoyl-sn-glycero-3-phospho-(1-rac-glycerol)(sodium salt),
$T_m=-2^\circ\text{C}$), and GM1 (ovine brain ganglioside GM1) were obtained from
Avanti Polar Lipids, Inc. (Alabaster, AL) in powder form and used without further
purification. Lipids were either dissolved in chloroform (DPPC and POPG) or 9/1 (v/v) chloroform/methanol (DMPE and GM1) to make
5 mg/mL stock solutions. Solvents were HPLC grade and obtained from Fisher Scientific (Pittsburgh, PA). SP-B 9-25 is a truncated synthetic peptide  of the 79-residue lung surfactant protein B. The peptide is an amphipathic helix with an amino acid sequence of WLCRALIKRIQAMIPKG. SP-B 9-25 was
prepared using Fmoc chemistry \cite{fields1991hbtu, waring2005role} (double coupling, 2 $\times$45 min) with reagents from Applied Biosystems (Foster City,CA) with a 431A solid phase peptide synthesizer (Applied Biosystems, Foster City, CA) or a Symphony/Multiplex SPPS synthesizer (Protein Technologies, Tucson, AZ). The peptide was cleaved using modified Reagent K (90$\%$ trifluoroacetic acid, 4.4$\%$ triisopropyl silane, 2.2$\%$ thioanisol, and 4.4$\%$ water). The peptide was precipitated upon the addition of ice-cold diethyl ether, collected by centrifugation, and purified by reversed-phase HPLC. MALDI-TOF mass spectrometry using an ABI Voyager RP-RBT2 reflection time-of-flight mass spectrometer (Applied Biosystems, Foster City, CA) confirmed the predicted molecular mass of the peptide. Lipid and lipid:peptide mixtures were prepared in several molar ratios: pure DMPE, DMPE:Chol at 5$\%$ and 10$\%$, DPPC:POPG 7:3, DPPC:POPG:SP-B 9-25 70:30:2, DPPC:GM1 8:2, and DPPC:GM1 5:5. The solutions were diluted with chloroform to obtain spreading solutions of concentration 0.1 mg/mL. Visualization with fluorescence microscopy was possible upon addition of 0.5 mol$\%$ Texas Red 1,2-dihexadecanoyl-sn-glycerol-3-phosphoethanolamine, triethylammonium salt (TR-DHPE) (Invitrogen, Carlsbad, CA) to the
spreading solutions. The subphase for all the experiments was ultrapure water (resistivity $\geq 18 M\Omega\,cm$) made using a ultrapurification (Milli-Q Advantage A10, Millipore, Bedford, MA). The superphase was air except for experiments with monolayers containing POPG. To minimize oxidative
damage to the unsaturated oleoyl chain of POPG, ultrahigh purity Argon 5.0 (Airgas, Chicago, IL) was used as the superphase.

\subsubsection{Instrument Setup}
All monolayer experiments are performed using a home-built Langmuir trough system\cite{Gopal}.  The set-up consists of a Teflon trough $(27.5cm\times6.25cm\times 0.63cm)$ fitted with two symmetrically mobile barriers $(l=6.25cm)$ made of the same material.  The barriers were placed on linear translational stages (UTM100CC, Corp., Irvine, CA) and interfaced to a Newport MM2000 motion controller to obtain movements with micron precision. The entire Langmuir trough also sits on x, y, and z translations stages (Newport, Irvine, CA) that allow for scanning of the air-water interface in each of those respective directions. The surface pressure measurements are made using a Wilhelmy surface balance (Reigler and Krstein, Berlin, Germany). As the surface area is reduced (compression) or increased (expansion) the change in surface pressure is monitored, giving rise to surface pressure $\pi [mN/m]$ versus area $A\,[\mathring{A}^2/molecule]$ isotherms.  Temperature of the water subphase is maintained within $0.1\,^o$C of the target temperature using a home built control assembly consisting of thermoelectric units (Omega Engineering INc, Stamford, CT) attached to a heat skin maintained at $20\,^o$C  by a Neslab RTE-100 water circulator (Portsmouth, NH). The trough is kept covered with a resistively heated indium tin oxide coated glass plate (Delta Technologies, Dallas, TX), which is maintained at approximately $2\,^o $C above the target subphase temperature in order to minimize air currents, reduce evaporative losses, and prevent condensation of water on the microscope objective. Direct imaging of the surface is performed using a fluorescence microscope with either a $50 \times$ or $20 \times$ extra-long working distance objective lens (Nikon Y-FL, Fryer Co., Huntley, IL). The filter cube (Nikon HYQ Texas Red, Fryer Co., Huntley, IL) used permits excitation between $530$ and $590 nm$ and emission between $610$ and $690 nm$. A CCD camera (Stanford Photonics , Inc., Palo Alto, CA) is used to collect images at 30 frames/s, which are recorded on a miniDV digital video cassette recorder (SONY, Tokyo, Japan).  The entire trough/microscope assembly is mounted on a vibration isolation table (Newport, Irvine, CA) and controlled completely by a custom software interface written using LabView 4.1 (National Instruments, Dallas, TX). 

\subsubsection{Isothermal compression measurements}\label{sec:meas}
For each monolayer experiment, 80 mL of water was poured into the trough. The subphase was heated to 25, 32, or 37$^\circ$C in the case of DPPC:POPG, 25$^\circ$C for DPPC:POPG:SP-B 9-25, 30$^\circ$C for DPPC: GM1, and 15, 25, or 37$^\circ$C for DMPE. When the target temperature was reached, the surface balance was calibrated to the value of surface tension of pure water for that temperature \cite{vargaftik1983international}. The monolayer was then spread at the gas/water interface by gently depositing drops onto the surface and allowed to equilibrate for 20 min. The barrier
compression was started (linear speed of 0.1 mm/s) and isotherm data were collected at 1-s intervals. The surface was imaged continuously throughout the compression.

\subsection{Nonlinear elastic response of lipid monolayers in the condensed phase}
\textcolor{black}{In the fully two-dimensional framework}, the HM law has been used for expressing the hydrostatic pressure $\Pi=(\sigma_1+\sigma_2)/2$ as a function of the deformation $\lambda$. Uniaxial strain assumptions also imply that the area change $J_G=\lambda$, so that isotherm curves connecting surface pressure and molecular area change can be re-expressed as stress-stretch relations. For convenience, the isotherm pressure data have been normalized by \textcolor{black}{considering a nominal value of} monolayer thickness $h$ of 2.5nm (i.e. $\Pi=\pi/h$) to give traditional pressure and moduli units by dimensional analysis.
Experimental data have been fitted by starting from the LE-C transition, assumed as the point in which the monolayer system starts to condensate by exhibiting a membrane-elasticity dominated response with a solid-like behaviour. This LE–C phase transition is marked by a drop in the experimental bulk modulus. It is tempting to quantitatively interpret this from a mechanistic standpoint of phase transitions and lipid condensation occurring at smaller scales\cite{carotenuto2021multiscale}. This implied to single out the local and global maxima of the experimental bulk modulus for selecting data points in the fitting procedures. \textcolor{black}{Data range for fitting has been determined by means of a custom-made detection algorithm, by performing the analysis of experimental bulk modulus data in order to detect the initial and ending points and by checking the fitting error with respect to a given accuracy.}\\
Chemically different lipid systems have been then considered by varying either temperature or composition. Starting from using hyperfoam model to follow the solid response of the monolayers, here the focus is on examining and characterizing theoretically the potential occurrence of a shear-banding instability, in such a case determining the corresponding banding pressure at the bifurcation.

\subsubsection{Observation of shear banding in tested lipid systems}
\begin{figure}[htbp] 
\centering
\includegraphics[width=.99
 \textwidth]{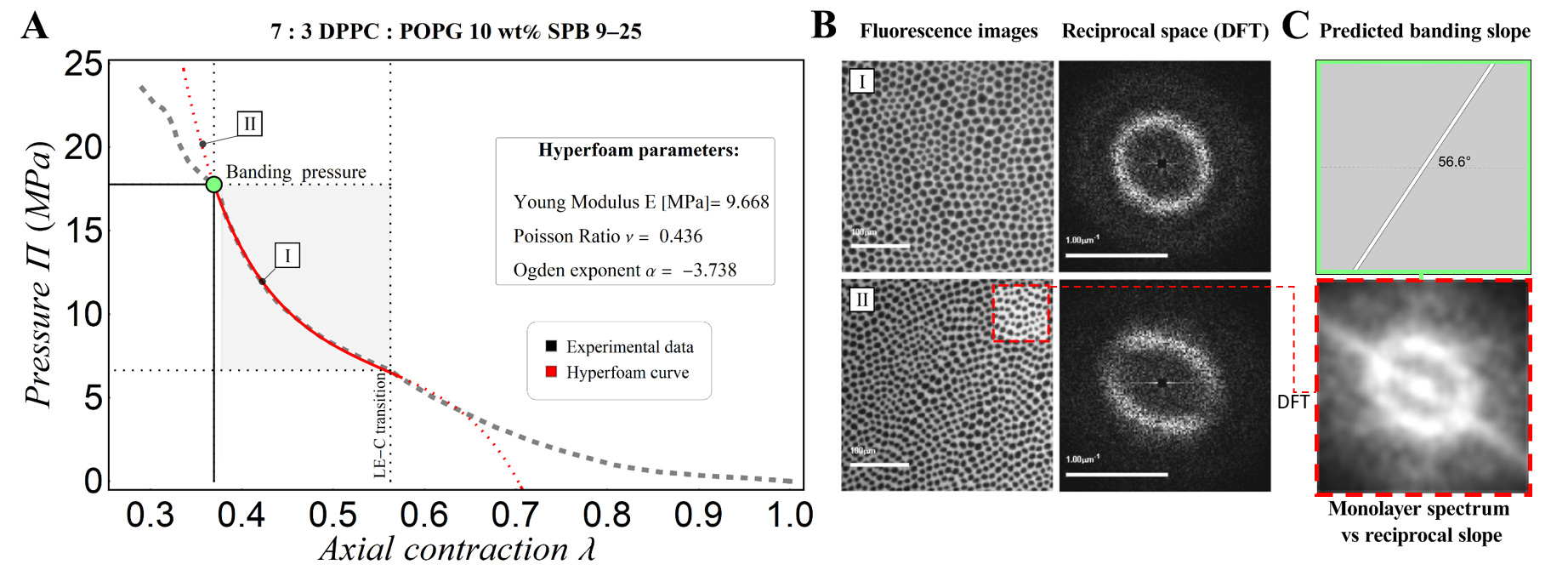}
\caption{ Estimation of shear banding in 7 : 3 DPPC : POPG 10 wt$\%$ SP-B 9 mixtures. \textbf{A} Experimental data with hyperfoam curve in the C-phase of the isotherm. Green disk marks the matching banding pressure, \textcolor{black}{while the gray square displays the adopted data range for fitting}. \textbf{B} Fluorescence images and relative Fourier transforms before and after banding, adapted from \cite{Pocivavsek1}. \textbf{C} Theoretically predicted banding slope reproducing the rotational rearrangement of lipid domains highlighted by means of in-frequency analysis.}
\label{fig.DPPC_fourier}
\end{figure}

The exposed methodology has been applied to different monolayer experiments. One of the systems under study is a model lung surfactant composed of dipalmitoylphosphatidylcholine (DPPC), palmitoyloleoylphosphatidylglycerol (POPG), and a surfactant peptide. The collapse and phase behaviour of the pure lipid system 7 : 3 DPPC : POPG has been extensively studied experimentally, by also providing some first estimations based on linear elasticity and plate models aimed at analysing their buckling \cite{Gopal, Diamant1, Diamant2, gopal2006microscopic, Pocivavsek1}, while the development of shear bands observed in some of these systems has not been yet theoretically investigated. To this aim, the macroscopic response of this monolayer in the condensed phase has been characterized by analysing the influence both of temperature and of the addition of a truncated lung surfactant peptide B on its material properties and its collapse behaviour. 
In particular, shear banding has been experimentally observed in DPPC : POPG with 10 wt$\%$ SP-B 9 lipid–peptide mixtures \textcolor{black}{from \cite{Pocivavsek1}}. In particular, Figure \ref{fig.DPPC_fourier}\hyperref[{fig.DPPC_fourier}]{A} shows the hyperfoam curve with the relative fitting parameters in comparison with the obtained data points in the C-phase. Experiments show a shear banding instability at a pressure of about $18$MPa, which is correctly predicted by calculating the critical stretch satisfying the l.o.e. condition $\mathcal{L}(\lambda)=0$ and evaluating the corresponding pressure. In order to gain insights into the effective instability mode, we considered fluorescence images of this compression test (taken from \cite{Pocivavsek1}) and the corresponding spectra, calculated by means of a Discrete Fourier Transform (DFT). As shown in Figure \ref{fig.DPPC_fourier}\hyperref[{fig.DPPC_fourier}]{B}, before banding (frame I) the monolayer shows an isotropic spectrum. After the banding event (frame II), the in-plane rearrangement of compressed domains leads to the emergence of preferential orientations, the entire spectrum highlighting a loss of isotropy. In particular, the distortion of the spectrum provides a deformation with higher spatial frequencies occurring in the direction of compression, which combines with a second vertical frequency by producing an overall rotation of the domain, this indicating an affine inclination of the spatial components. On this bases, we compared the theoretical banding slope calculated by using equation \eqref{ratio} with the bands generated in a lipid representative area element (a $\delta\times\delta$ region  introduced in \cite{carotenuto2021multiscale}, with size of the order of tens of microns), by directly overlapping the respective Fourier transforms, see Figure \ref{fig.DPPC_fourier}\hyperref[{fig.DPPC_fourier}]{C}. From this comparison, it is apparent how the reciprocal banding line matches the major axis of the frame spectrum, this confirming the effectiveness in reproducing the actual re-orientation mechanism that produces macroscopic banding.

\begin{figure}[htbp] 
\centering
\includegraphics[width=.99
 \textwidth]{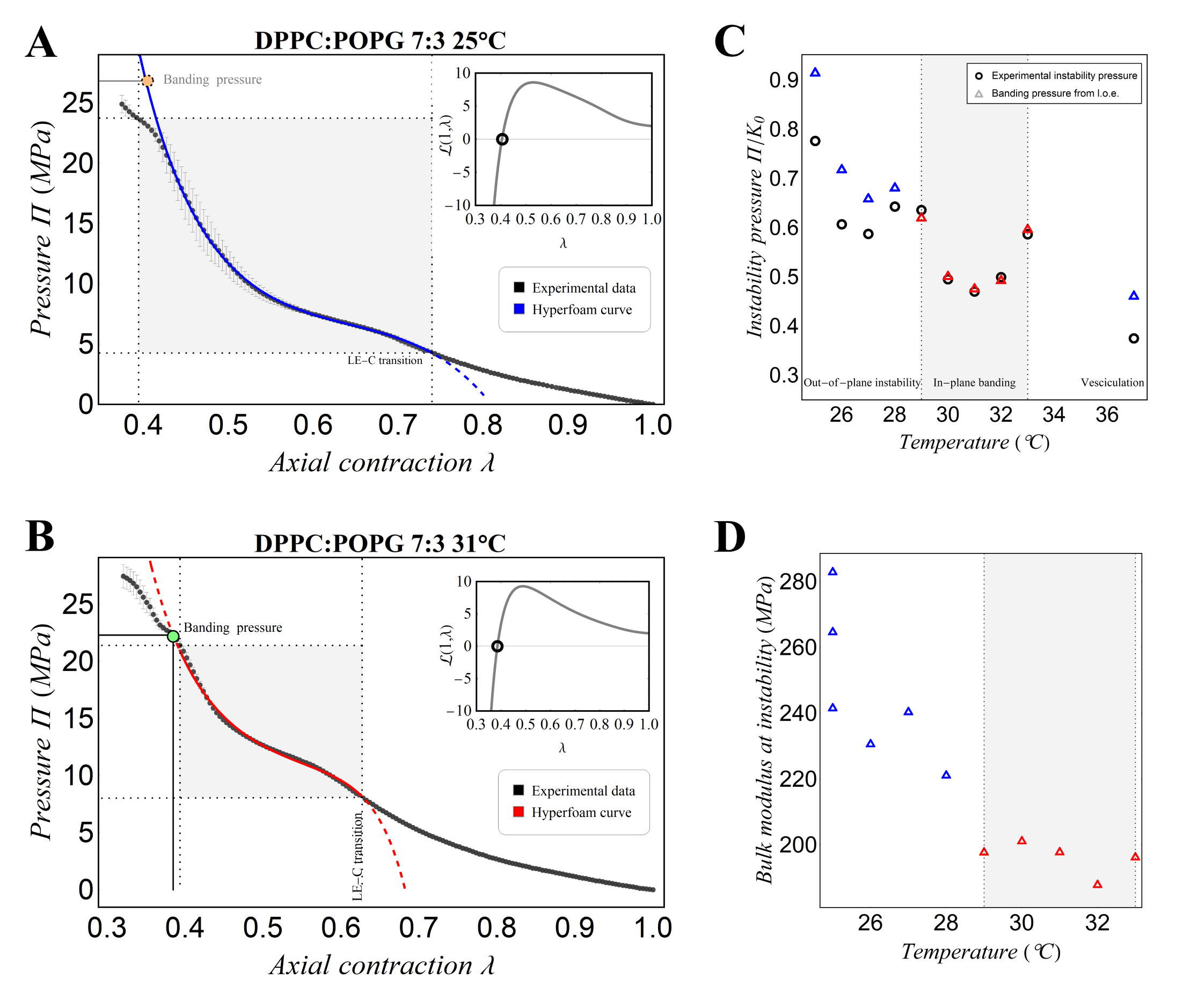}
\caption{Analysis of effective shear banding in 7 : 3 DPPC : POPG systems at varying temperature. \textbf{A} Experimental and theoretical curves at 25$^\circ$C showing inactive banding. \textbf{B} Experimental and theoretical curves at 31$^\circ$C showing effective banding. \textbf{C} Influence of temperature in triggering planar re-organization in lipid monolayers. \textbf{D} Variation of compaction (tangent) modulus with the temperature. \textcolor{black}{Red color is used to represent data for cases showing shear banding, while blue color refers to data with other collapse mechanisms.}}
\label{fig.DPPC_temp}
\end{figure}

\textcolor{black}{Besides literature data, the mechanical response of other 7 : 3 DPPC : POPG monolayers has been investigated at varying temperatures, by carrying out a campaign of experiments to obtain new isotherm compression curves between 25$^\circ$C and 37$^\circ$C (see Section \ref{sec:meas})}. By referring to the phase diagram of Figure \ref{fig.fig1}\hyperref[{fig.fig1}]{C}, one can see that 7 : 3 DPPC : POPG mixtures experimentally exhibit different collapse mechanisms, with a transition from folding to banding modes as temperature increases. In fact, out-of-plane instabilities have been mostly observed at lower temperatures, while in-plane strain localization starts to be detected at about $30^\circ$C for this type of monolayer. This behaviour can be well traced back by the theoretical analyses carried out on the different isotherm curves. By the way of example, Figures \ref{fig.DPPC_temp}\hyperref[{fig.DPPC_temp}]{A} and \hyperref[{fig.DPPC_temp}]{B} compare the curves obtained respectively at $25^\circ $C and $31^\circ $C with the corresponding fitting. \\As illustrated, in the former case (Figure \ref{fig.DPPC_temp}\hyperref[{fig.DPPC_temp}]{A}), the theoretical banding bifurcation occurs above the experimental instability point, in correspondence of which stiffness relaxation can be likely ascribed to the onset of early wrinkling modes (not detectable through the used imaging technique) that precede the actual folding collapse of the monolayer observed at pressures of about $29$MPa. \textcolor{black}{This is in accord with experimental evidences summarized in Figure \ref{fig.fig1}\hyperref[{fig.fig1}]{C}, where the 7 : 3 DPPC : POPG monolayers at $25^\circ $C are found to exhibit folding-type collapse modes.} On the other hand, at $31^\circ $C (Figure \ref{fig.DPPC_temp}\hyperref[{fig.DPPC_temp}]{B}), the theoretical banding pressure coincides with the actual bifurcation point, at which in-plane shear bands effectively emerge from morphological analyses. As also synthetically represented in Figure \ref{fig.DPPC_temp}\hyperref[{fig.DPPC_temp}]{C} (in which pressures are normalized by the in plane bulk modulus $K_0=(E_1+E_2)/2(1+\nu)(1-2\nu)$ at small strains), this particular lipid system can always exhibit a potential shear banding, but this mechanism is found to take place only in the temperature range $29^\circ \text{C}-32^\circ \text{C}$, where the theoretical and experimental instability pressures actually concur, in agreement with the outcomes of the compression tests. In addition, Figure \ref{fig.DPPC_temp}\hyperref[{fig.DPPC_temp}]{D} quantitatively confirms what already observed in Pocivavsek et al. \cite{Pocivavsek2}, reporting that monolayers with a higher bulk modulus in the compaction region generally collapse out-of-plane whereas more compressible monolayers relax the stress by showing in-plane instability at higher temperatures.

\begin{figure}[htbp] 
\centering
\includegraphics[width=.99
 \textwidth]{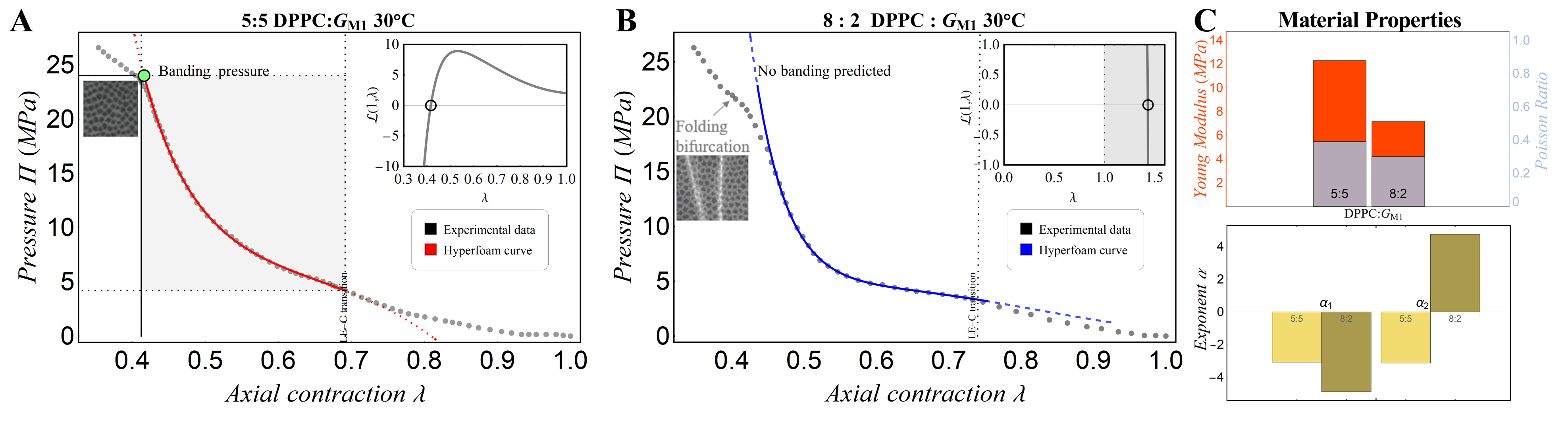}
\caption{Analysis of banding collapse in DPPC : GM1 monolayers at 30$^\circ$C. \textbf{A} 5 : 5 DPPC : GM1 banding monolayer and \textbf{B} 8 : 2 DPPC : GM1 folding monolayer. \textcolor{black}{Red color is used to represent data for cases showing shear banding, while blue color refers to data with other collapse mechanisms.} \textbf{C} Obtained hyperfoam material constants.}
\label{fig.DPPC_GM1}
\end{figure}

The isotherms of DPPC : GM1 mixtures at two different concentrations have been also analysed by exploiting test curves reported in \cite{Pocivavsek2}. Results reported in Figure \ref{fig.DPPC_GM1} show that the 5 : 5 DPPC : GM1 monolayer (Figure \ref{fig.DPPC_GM1}\hyperref[{fig.DPPC_GM1}]{A}) undergoes shear banding at a pressure of approximately 24MPa. Differently,  the 8 : 2 DPPC : GM1 mixture in Figure \ref{fig.DPPC_GM1}\hyperref[{fig.DPPC_GM1}]{B} manifested an out-of-plane collapse by forming folds, as also shown in the inset. Accordingly, the theoretical HM stress-stretch curve retraced the isotherm points by not incurring into a loss of ellipticity in contraction (as highlighted in the inset, the condition $\mathcal{L}=0$ in fact occurred at $\lambda\approx 1.5$, a not acceptable value for the case at hand). In terms of mechanical properties (see Figure \ref{fig.DPPC_GM1}\hyperref[{fig.DPPC_GM1}]{C}), the banding monolayer starts with a higher initial Young modulus and gradually evolves with higher slope during all the transient stage of nucleation, while the 8:2 folding monolayer shows a quasi-horizontal initial phase and then hardens at densification during the last branch of the condensed phase.\\ \\

\begin{figure}[htbp] 
\centering
\includegraphics[width=.35
 \textwidth]{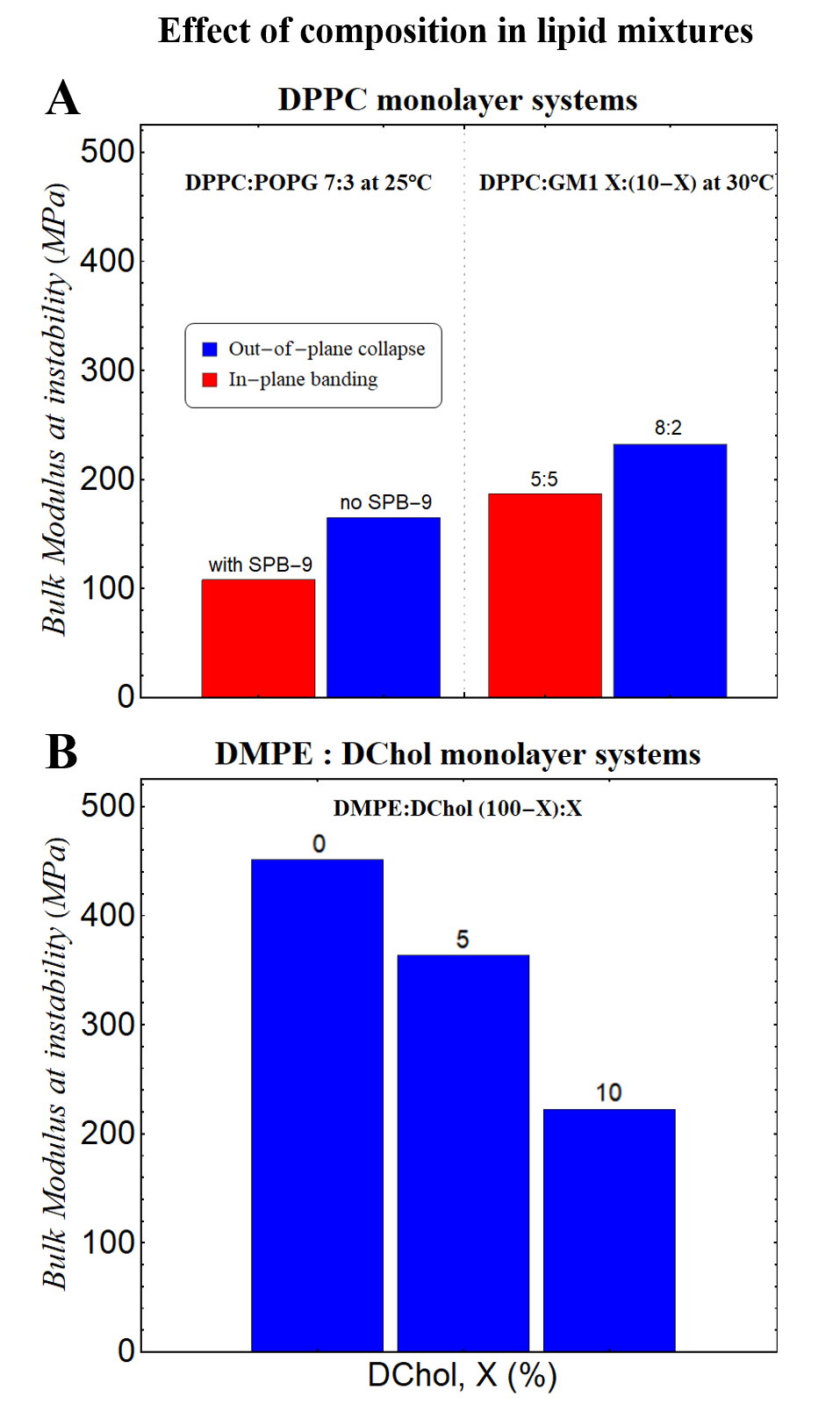}
\caption{\textcolor{black}{Effect of chemical composition on lipid mixtures in terms of compaction moduli at the instability onset. \textbf{A}. Analyses of i) 7 : 3 DPPC : POPG 10 wt$\%$ SP-B 9 versus 7 : 3 DPPC : POPG at 25$^\circ$C and ii) 5:5 DPPC : GM1 versus 8:2 DPPC : GM1 at 30$^\circ$C. \textbf{B}. Bulk moduli of DMPE systems, pure lipid monolayer versus 5$\%$ and 10$\%$ DChol mixtures.}}
\label{fig.composition}
\end{figure}

\textcolor{black}{
To thus highlight the effect of composition on monolayers' mechanical response, the bulk moduli at the instability onset in the compaction phase have been analysed for chemically different lipid systems and synoptically represented in Figure \ref{fig.composition} (moduli estimation were performed according to the expression reported in \hyperref[{sec:appendix}]{Appendix} and with reference to the material parameters found in Table \ref{Table}). In particular, the direct comparison of 7 : 3 DPPC : POPG at 25$^\circ$C mixed with SPB-9 and as a pure lipid system --respectively illustrated in the tests of Figure \ref{fig.DPPC_fourier}\hyperref[{fig.DPPC_fourier}]{A} and Figure \ref{fig.DPPC_temp}\hyperref[{fig.DPPC_temp}]{A}-- has been first carried out. As shown in  Figure \ref{fig.composition}\hyperref[{fig.composition}]{A}, the bulk modulus at instability for the lipids without the peptide results to be approximately 165 MPa, while decreases to about 110 for DPPC:POPG:SP-B 9, which undergoes banding.
Figure \ref{fig.composition}\hyperref[{fig.composition}]{A} also shows an analogous effect for DPPC : GM1 mixtures reported in Figure \ref{fig.DPPC_GM1}\hyperref[{fig.DPPC_GM1}]{A} and \hyperref[{fig.DPPC_GM1}]{B}. In this case, the ultimate compaction modulus of 5 : 5 DPPC : GM1 monolayer of about 190 MPa is compared to the stiffer 8 : 2 DPPC : GM1 mixture, whose bulk modulus in the densification zone approaches about 250 MPa.
Additionally, Figure \ref{fig.composition}\hyperref[{fig.composition}]{B} reports the effect of dihydrocholesterol  (DChol) in DMPE:DChol mixtures. Also in this case, the compaction modulus of the mixtures tends to increase in systems for higher lipid contents, being it maximum in the control lipid system.
}

\begin{figure}[htbp] 
\centering
\includegraphics[width=.99
 \textwidth]{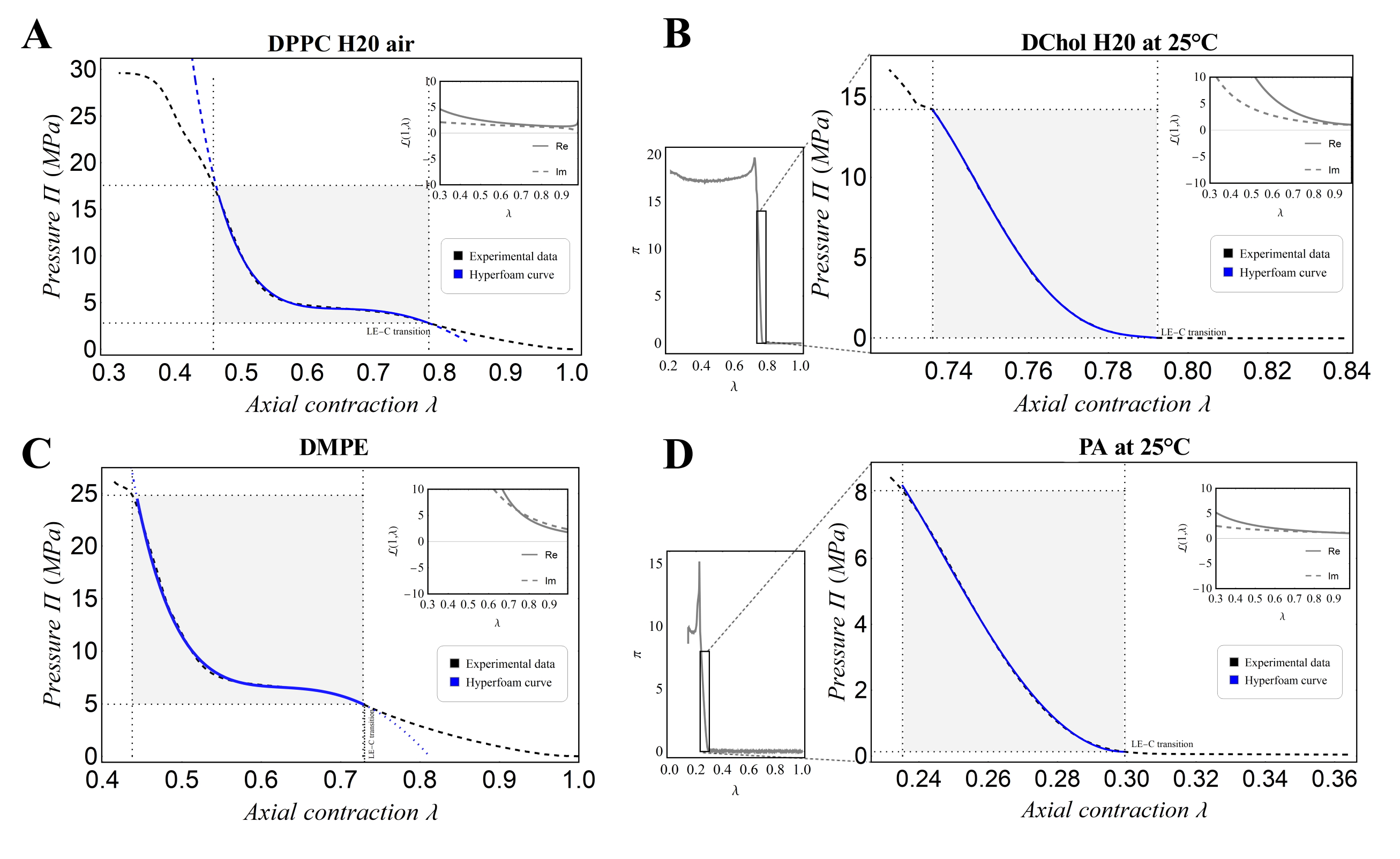}
\caption{Solid response of some traditional pure monolayers that avoid shear banding collapse: \textbf{A} DPPC, \textbf{B} Dchol, \textbf{C} DMPE and \textbf{D} \textcolor{black}{Palmitic acid (16:0, PA)} systems.}
\label{fig.Pure}
\end{figure}

This feature seems to be preserved in some pure systems of canonical use in Langmuir experiments, which in the most of cases exhibit a sharper transition from gaseous to solid elastic phase, there showing abrupt compaction. All these monolayers collapse by buckling and in-plane instabilities have been not observed, this motivating to therefore consider them as a control for the proposed methodology. As reported in Figure \ref{fig.Pure}, the different systems of pure lipids provided exhibit a knee-shaped response, for which the associated l.o.e. conditions assume complex values by not admitting a critical stretch, so well reproducing the experimental evidences. All the obtained fitting parameters for the different cases analysed are reported in Table \ref{Table}.

\section{Conclusions and perspectives}
The mechanisms of monolayer collapse are highly investigated phenomena for their importance into the understanding of the mechano-biological role of lipid membranes in many living structures. To the best of Authors' knowledge, the vast majority of literature works mainly focused on out-of-plane instability by analysing buckling and wrinkling-to-folding processes through both experiments on Langmuir monolayers and linearly elastic mechanical models. Nevertheless, strong evidence of the existence of in-plane collapses with rearrangement of condensed domains --recognized as shear bands-- are also documented, which however lack both of a theoretical interpretation and of a mathematical prediction.\\
By then starting from the largely accepted assumption that lipid monolayers behave as elastic systems in the solid-like C-phase and by observing that therein they show a nonlinear response, we have hypothesized that the variety of the features exhibited by pressure-area change curves could be traced back by adopting a hyperfoam model. This assumption lays on similarities between the macroscopic responses of lipid material and cellular solids, which both provide an initial transient phase with internal rearrangement mechanisms (possibly captured by means of Structured Deformations) followed by a densification branch.
This particular constitutive assumption was proposed as a possible way to model monolayer response from LE-C transition to collapse, so enriching the well-established equation of state commonly adopted to represent isotherms in the gaseous and liquid expanded phases. In addition, after deriving the general conditions governing the loss of ellipticity of tangent stiffness tensor by studying the incremental mechanical problem, the obtained hyperfoam responses were exploited to theoretically evaluate the potential occurrence of shear banding instabilities through which lipid monolayers relax stress by in-plane localizing deformation along preferentially oriented lines. The critical stretch and pressure was accordingly predicted, by validating the obtained banding slopes through direct imaging techniques.\\
The proposed approach can therefore suggest a method to model the monolayer response in the framework of nonlinear elasticity and it could be certainly enriched by including different aspects. A first theoretical improvement could be the direct modelling of the lipid clusters re-arrangements occurring at lower scales through enhanced geometric descriptions of membrane kinematics based on the Structured Deformation theory \cite{carotenuto2020raft,carotenuto2021multiscale}. Here this is  introduced only in a formal setting of the problem by considering the potential influence of possible inelastic deformations related to lipid conformational changes that would hence contribute to the total strain. This could let to implement coupled mechanical models explicitly including the effect of lipid cluster condensation. Other features could be analysed as well, such as hysteresis (possibly related to the inelastic effects due to the submacroscopic rearrangements) and visco-elasticity effects. \textcolor{black}{The mechanism rate, either in terms of the effect of the loading rate or of the imposed deformation rate producing creep and/or stress relaxation, could be discussed by means of a suitable nonlinear viscoelastic characterization (for instance through a Doi-Edwards-like approach \cite{marrucci1986doi, ianniruberto2017shear}).}
The influence of non-uniform pre-stress, e.g. coming from the aforementioned multi-scale models, could be examined in order to investigate more complicated banding mechanisms through the use of ad hoc numerical algorithms. The presented methodology could also be applied to characterize out-of-plane buckling, by deriving critical loads as a function of the pre-stress accumulated under large compression. 
Instabilities of lipid monolayers seem to be associated to the development and/or to the triggering of ARDS, and it is then felt that prediction of banding/wrinkling in such systems could be used as early targeting or even as a biomechanical marker for such diseases.

\section*{Acknowledgements} 
A.R.C. acknowledges the support of the grant no. PON-AIM1849854-1 from the Italian Ministry of Education, University and Research (MUR).
L.D. and M.F. acknowledge the partial support from the grants: (i)   Italian Ministry of Research (MIUR) through the grants PRIN-20177TTP3S and (ii) PON
“Stream”-ARS01 01182. L.D. also acknowledges support from the ERC through (iii) FET Open “Boheme” grant no. 863179, (iv) LIFE GREEN VULCAN LIFE19 ENV/IT/000213, and (v) ERC-ADG-2021-101052956-BEYOND. KYCL acknowledges support from The University of Chicago Materials Research Science and Engineering Center (NSF/DMR- 2011854). This work was also partially supported by the National Science Foundation through MCB-1950525 (to KYCL).

\newpage
\section*{Appendix}\label{sec:appendix}
\subsection*{Expressions for loss of ellipticity conditions}
\setcounter{equation}{0}
\renewcommand{\theequation}{A.\arabic{equation}}

In case of a HM with a single-mode, i.e. $n=1$, the l.o.e. condition turns out to be:

\begin{equation}\label{loe1}
\resizebox{\textwidth}{!}
{$
\mathcal{L}=\frac{\bar{\nu } \left(G_1 G_2\right){}^{\frac{\alpha  \nu }{1-2 \nu }} \left(\alpha ^- G_1^{\alpha }
   G_2^{\alpha } \left(G_1^2 \alpha ^--G_2^2 \left(\alpha ^-+2 \mathcal{R}\right)\right)+G_2^2 G_1^{2
   \alpha } \left(2 \mathcal{R} \alpha ^-+1\right)-G_1^2 G_2^{2 \alpha }\right)+\left(\nu  \alpha
   ^=+1\right) \left(-\alpha ^-\,G_1^{\alpha +2}+G_2^{\alpha +2} \left(\alpha ^-+2
   \mathcal{R}\right)-G_1^2 \alpha ^+ G_2^{\alpha }+G_2^2 G_1^{\alpha } \left(\alpha ^+-2
   \mathcal{R}\right)\right)}{2 G_2^2 \left(G_1^{\alpha }-G_2^{\alpha }\right) \left(\bar{\nu } \alpha ^-
   G_1^{\alpha } \left(G_1 G_2\right){}^{\frac{\alpha  \nu }{1-2 \nu }}-\nu  \alpha ^=-1\right)}
$}
\end{equation}
 
where the positions $\bar{\nu}=2\nu-1$, $\alpha^\pm=\alpha\pm 1$ and $\alpha^{=}=\alpha-2$ have been introduced, while the $\mathcal{R}$ root factor reads as:
\begin{equation}
\mathcal{R}=\sqrt{\frac{G_1^2 \left(\bar{\nu } \alpha ^- G_2^{\alpha } \left(G_1 G_2\right){}^{\frac{\alpha  \nu }{1-2
   \nu }}+\bar{\nu }-\alpha  \nu \right)}{G_2^2 \left(\bar{\nu } \alpha ^- G_1^{\alpha } \left(G_1
   G_2\right){}^{\frac{\alpha  \nu }{1-2 \nu }}+\bar{\nu }-\alpha  \nu \right)}}
\end{equation}
which explicitly depends on $\alpha$ and $\nu$. For an HM law with $n=2$, l.o.e. condition has a lengthy expression 

\begin{equation}
\resizebox{\textwidth}{!}
{$
\mathcal{L}=\frac{
-\alpha _2 \left(\nu  \alpha _1^=+1\right) \left(G_1 G_2\right){}^{\frac{\alpha _1 \nu }{\bar{\nu}}}\,A
+ \alpha _1 \alpha _2 \gamma  \left(G_1 G_2\right){}^{\frac{\alpha _2 \nu }{\bar{\nu }}}\, (2 G_1 \mathcal{R}_2 G_2^{\alpha _1+1} \left(\nu  \alpha _2^=-\alpha _2 \nu +1\right)-\nu\,A) + \bar{\nu }\, B
}{2 G_2^2 \left(\alpha _1 \gamma  G_1^{\alpha _2}-\alpha _1 \gamma  G_2^{\alpha _2}+\alpha _2 G_1^{\alpha
   _1}-\alpha _2 G_2^{\alpha _1}\right) \left(\bar{\nu } \left(\alpha _1 \gamma  \left(G_1
   G_2\right){}^{\frac{\alpha _2 \nu }{\bar{\nu }}}+\alpha _2 G_1^{\alpha _1} \alpha _1^-+\alpha _1 \gamma
    G_1^{\alpha _2} \alpha _2^-\right)+\alpha _2 \left(-\left(\nu  \alpha _1^=+1\right)\right) \left(G_1
   G_2\right){}^{\frac{\alpha _1 \nu }{\bar{\nu }}}-\alpha _1 \alpha _2 \gamma  \nu  \left(G_1
   G_2\right){}^{\frac{\alpha _2 \nu }{\bar{\nu }}}\right)}
   $}
\end{equation}

where 
\begin{equation}
\begin{split}
A=&-G_2^2 \left(\alpha _2 G_2^{\alpha _1} \alpha _1^-+\alpha _1 \gamma  G_2^{\alpha _2} \alpha
   _2^-\right)+\alpha _2 G_1^{\alpha _1+2} \alpha _1^-+\alpha _1 \gamma  G_1^{\alpha _2+2} \alpha_2^- -\alpha_1 \gamma  G_2^2 \alpha _2^+ G_1^{\alpha _2}+\\
 &+G_1^2 \left(\alpha _1 \gamma  \alpha_2^+
   G_2^{\alpha _2}+\alpha _2 \alpha _1^+ G_2^{\alpha _1}\right)-\alpha _2 G_2^2 \alpha_1^+ G_1^{\alpha
   _1}+2 \alpha _1 \gamma  G_2 \mathcal{R}_2 G_1^{\alpha _2^+}+2 \alpha _2 G_2 \mathcal{R}_2 G_1^{\alpha
   _1^+}+\\
&-2 G_2 G_1 \mathcal{R}_2 \left(\alpha _1 \gamma  G_2^{\alpha _2}+\alpha _2 G_2^{\alpha _1}\right),
\end{split}
\end{equation}

and 

\begin{equation}
\begin{split}
B&=\alpha _2^2 G_1^{\alpha _1+2} G_2^{\alpha _1} \alpha _1^{2 -}-\alpha _2^2 G_1^{\alpha _1} G_2^{\alpha _1+2} \alpha _1^{2 -}-\alpha _2^2 G_1^2
   G_2^{2 \alpha _1}+\alpha _2^2 G_2^2 G_1^{2 \alpha _1}+\\
&+\gamma  \left(\alpha _1 \alpha _2 G_1^{\alpha _1+2} \alpha _1^- \left(G_1 G_2\right){}^{\frac{\alpha _2 \nu }{\bar{\nu }}}-2 \alpha _1 \alpha _2
   G_1^2 G_2^{\alpha _1+\alpha _2}+2 \alpha _1 \alpha _2 G_2^2 G_1^{\alpha _1+\alpha _2}\right.+\\
&-\alpha _1 \alpha _2 G_2^{\alpha _1+2} \alpha _1^- \left(G_1 G_2\right){}^{\frac{\alpha _2 \nu }{\bar{\nu }}}+\alpha _1 \alpha _2
   G_1^2 G_2^{\alpha _1} \alpha _1^+ \left(G_1 G_2\right){}^{\frac{\alpha _2 \nu }{\bar{\nu }}}-\alpha _1 \alpha _2 G_2^2 G_1^{\alpha _1} \alpha
   _1^+ \left(G_1 G_2\right){}^{\frac{\alpha _2 \nu }{\bar{\nu }}}+\\
&\left.+\alpha _1 \alpha _2 G_1^{\alpha _2+2} G_2^{\alpha _1} \alpha _1^- \alpha _2^--\alpha _1 \alpha _2 G_1^{\alpha _2} G_2^{\alpha _1+2} \alpha _1^-
   \alpha _2^-+\alpha _1 \alpha _2 G_1^{\alpha _1+2} G_2^{\alpha _2} \alpha _1^- \alpha _2^--\alpha _1 \alpha _2 G_1^{\alpha _1} G_2^{\alpha _2+2}
   \alpha _1^- \alpha _2^-\right)+\\
&+\gamma ^2 \left(\alpha _1^2 \left(G_1^{\alpha _2+2}-G_2^{\alpha _2+2}\right) \alpha _2^- \left(G_1 G_2\right){}^{\frac{\alpha _2 \nu }{\bar{\nu
   }}}+\alpha _1^2 G_1^2 \alpha _2^+ G_2^{\alpha _2} \left(G_1 G_2\right){}^{\frac{\alpha _2 \nu }{\bar{\nu }}}-\alpha _1^2 G_2^2 \alpha _2^+
   G_1^{\alpha _2} \left(G_1 G_2\right){}^{\frac{\alpha _2 \nu }{\bar{\nu }}}+\right.\\
&+\left.\alpha _1^2 G_1^{\alpha _2+2} G_2^{\alpha _2} \alpha _2^{2 -}-\alpha
   _1^2 G_1^{\alpha _2} G_2^{\alpha _2+2} \alpha _2^{2 -}-\alpha _1^2 G_1^2 G_2^{2 \alpha _2}+\alpha _1^2 G_2^2 G_1^{2 \alpha _2}\right)+\\
&+\mathcal{R}_2 \left(2 \alpha _1^2 \gamma ^2 G_2 G_1^{\alpha _2^+} \left(G_1 G_2\right){}^{\frac{\alpha _2 \nu }{\bar{\nu }}}+2 \alpha _1 \alpha _2
   \gamma  G_2 G_1^{\alpha _1^+} \left(G_1 G_2\right){}^{\frac{\alpha _2 \nu }{\bar{\nu }}}\right.+\\
&-2 \alpha _1^2 \gamma ^2 G_1 G_2^{\alpha_2+1}
   \left(G_1 G_2\right){}^{\frac{\alpha _2 \nu }{\bar{\nu }}}-2 \alpha _1 \alpha _2 \gamma  G_2^{\alpha _2+1} \alpha _1^- G_1^{\alpha _1^+}-2
   \alpha _1 \alpha _2 \gamma  G_2^{\alpha _1+1} \alpha _2^- G_1^{\alpha _2^+}+\\
&+2 \alpha _2^2 G_2 G_1^{2 \alpha _1+1} \alpha _1^--2 \alpha _2^2
   \left(G_1 G_2\right){}^{\alpha _1+1} \alpha _1^-+2 \alpha _1^2 \gamma ^2 G_2 G_1^{2 \alpha _2+1} \alpha _2^-+\\
&\left.-2 \alpha _1^2 \gamma ^2 \left(G_1
   G_2\right){}^{\alpha _2+1} \alpha _2^-+2 \alpha _1 \alpha _2 \gamma  G_2 \left(\alpha _2^=+\alpha _1\right) G_1^{\alpha _1^++\alpha _2}\right)
\end{split}
\end{equation}
where analogous positions 
$\alpha_i^\pm=\alpha_i\pm 1$ and $\alpha^{=}=\alpha_i-2$ have been introduced, while 

\begin{equation}
\resizebox{\textwidth}{!}
{$
\mathcal{R}_2=\sqrt{\frac{\alpha _2 \left(\alpha _1 \gamma  \nu  \left(G_1 G_2\right){}^{\frac{\alpha _2 \nu }{\bar{\nu }}}+\left(\alpha _1 \nu -2 \nu +1\right)
   \left(G_1 G_2\right){}^{\frac{\alpha _1 \nu }{\bar{\nu }}}\right)-\bar{\nu } \left(\alpha _1 \gamma  \left(\left(G_1 G_2\right){}^{\frac{\alpha
   _2 \nu }{\bar{\nu }}}+G_2^{\alpha _2} \alpha _2^-\right)+\alpha _2 G_2^{\alpha _1} \alpha _1^-\right)}{\alpha _2 \left(\alpha _1 \gamma  \nu 
   \left(G_1 G_2\right){}^{\frac{\alpha _2 \nu }{\bar{\nu }}}+\left(\alpha _1 \nu -2 \nu +1\right) \left(G_1 G_2\right){}^{\frac{\alpha _1 \nu
   }{\bar{\nu }}}\right)-\bar{\nu } \left(\alpha _1 \gamma  \left(\left(G_1 G_2\right){}^{\frac{\alpha _2 \nu }{\bar{\nu }}}+G_1^{\alpha _2}
   \alpha _2^-\right)+\alpha _2 G_1^{\alpha _1} \alpha _1^-\right)}}
$}
\end{equation}

\subsubsection*{Expression of compaction modulus}
The compaction modulus $K_t$ exhibited during the densification phase can be calculated starting from the linearization of the pressure $\Pi(G_1,G_2)$ in correspondence of a generic deformation state $\left(\tilde{G}_1,\tilde{G}_2\right)$, 

\begin{equation}
\Pi\simeq \left.\Pi,_{G_1}\right|_{\left(\tilde{G}_1,\tilde{G}_2\right)}\,(G1-\tilde{G}_1)+\left.\Pi,_{G_2}\right|_{\left(\tilde{G}_1,\tilde{G}_2\right)}\,(G2-\tilde{G}_2)
\end{equation}

In case of uniaxial stretch, i.e. $G_2\equiv 1$ and $n=2$, the modulus $K_t$ is given by the slope of the pressure curve and results to be 

\begin{equation}
\resizebox{\textwidth}{!}
{$
K_t(\tilde{G}_1,1)=\frac{\alpha _2 E_1 \left(\left(\alpha
   _1-1\right) (2 \nu -1) \tilde{G}_1^{\alpha
   _1}-2 \left(\left(\alpha _1-2\right) \nu
   +1\right) \tilde{G}_1^{\frac{\alpha _1 \nu
   }{2 \nu -1}}-2 \nu +1\right)+\alpha _1 E_2
   \left(\left(\alpha _2-1\right) (2 \nu -1)
   \tilde{G}_1^{\alpha _2}-2
   \left(\left(\alpha _2-2\right) \nu
   +1\right) \tilde{G}_1^{\frac{\alpha _2 \nu
   }{2 \nu -1}}-2 \nu +1\right)}{2 \alpha _1
   \alpha _2 (\nu +1) (2 \nu -1)
   \tilde{G}_1^2}
   $}
\end{equation}

where $\tilde{G}_1$ on isotherm is the  axial contraction $\lambda$ at which modulus is calculated normalized by the deformation at which LE-C transition starts to occur, i.e. $\tilde{G}_1=\lambda/\lambda_{LE-C}$, the particular case $\tilde{G}_1=1$ returning the initial tangent modulus $K_0$.

\subsection*{Table of fitting parameters}

\begin{table}[h]
\centering
\resizebox{0.95\textwidth}{!}{
\begin{tabular}{|c|c|c|c|c|c|c|}
\hline
\rowcolor[HTML]{EFEFEF} 
\multicolumn{1}{|c|}{\cellcolor[HTML]{EFEFEF}Lipid system} & Temperature ($^\circ$C) & $E_1$ (MPa) & $E_2$ (MPa) & $\nu$ & $\alpha_1$ & \multicolumn{1}{c|}{\cellcolor[HTML]{EFEFEF}$\alpha_2$} \\ \hline
7 : 3 DPPC : POPG 10 wt$\%$ SP-B 9                         & 25                     & 9.668       & 9.668       & 0.436 & -3.738     & -3.738                                                  \\ \hline

& 25                     & 3941.179    & -3929.292   & 0.427 & -2.597     & -2.590                                                  \\
                                                           & 26                     & 501.049     & -490.225    & 0.441 & -4.424     & -4.423                                                  \\
                                                           & 27                     & 185.376     & -174.578    & 0.449 & -4.431     & -4.431                                                  \\
                                                           & 28                     & 106.622     & -94.862     & 0.437 & -4.687     & -4.687                                                  \\
                                                           & 29                     & 829.817     & -818.518    & 0.440 & -4.539     & -4.540                                                  \\
                                                          7 : 3 DPPC : POPG                                         
                                                           & 30                     & 18.091      & -6.320      & 0.451 & -4.525     & -4.525                                                  \\
                                                           & 31                     & 500.517     & -488.199    & 0.454 & -4.056     & -4.056                                                  \\
                                                           & 32                     & 74.432      & -61.824     & 0.445 & -4.712     & -4.712                                                  \\
                                                           & 33                     & 3.825       & 8.804       & 0.435 & -4.559     & -4.699                                                  \\
                                                           & 37                     & 668.037     & -656.806    & 0.464 & -3.035     & -3.050                                                  \\ \hline
5 : 5 DPPC : GM1                                           & 30                     & 507.197     & -494.902   & 0.391 & -3.125     & -3.071                                                  \\
8 : 2 DPPC : GM1                                           & 30                     & 77.881      & -70.716     & 0.299 & -4.882     & 4.756                                                   \\ \hline
DPPC                                                       & 25                     & 4166.325    & -4143.008   & 0.281 & 0.7068     & -0.0769                                                 \\
DChol                                                      & 25                     & 90148.941   & -90129.547  & 0.293 & 8.8372     & -8.2023                                                 \\
DMPE                                                       & 30                     & 1332.251    & -1314.546   & 0.375 & -2.0457    & -1.8671                                                 \\
PA                                                         & 25                     & 18573.7135  & -18573.7016 & 0.286 & -0.10629   & 0.51902                                                 \\ \hline 
DMPE : Dchol 95:5                                                         & 30                     & 942.355  & -927.912 & 0.365 & -4.089   & -3.759
\\                                          DMPE : Dchol 90:10                                                         & 30                     & 420.146  & -394.969 & 0.335 & -4.726   & -3.108
\\ \hline
\end{tabular}
}
\caption{Table of HM fitting parameters}
\label{Table}
\end{table}

\bibliographystyle{spphys}
\bibliography{bibmonolayer}

\newpage
\bibliographystyle{unsrt}  

\end{document}